\documentclass[letterpaper]{article} % DO NOT CHANGE THIS
\usepackage{aaai2026}  % DO NOT CHANGE THIS
\nocopyright
\usepackage{times}  % DO NOT CHANGE THIS
\usepackage{helvet}  % DO NOT CHANGE THIS
\usepackage{courier}  % DO NOT CHANGE THIS
\usepackage[hyphens]{url}  % DO NOT CHANGE THIS
\usepackage{graphicx} % DO NOT CHANGE THIS
\usepackage{natbib}  % DO NOT CHANGE THIS AND DO NOT ADD ANY OPTIONS TO IT
\usepackage{caption} % DO NOT CHANGE THIS AND DO NOT ADD ANY OPTIONS TO IT
\DeclareCaptionStyle{ruled}{labelfont=normalfont,labelsep=colon,strut=off} % DO NOT CHANGE THIS
\usepackage{amsmath}
\usepackage{amssymb}
\usepackage{bm}
\usepackage{booktabs}
\usepackage{array}
\usepackage{multirow}
\usepackage{xcolor}
\newcommand{\stableshare}{A}

\title{Chance, Persistent Advantage, and the Generative-AI Era in Open-Source Package Careers}
\author{Hazem Ibrahim\textsuperscript{\rm 1,$*$}, Yasir Zaki\textsuperscript{\rm 1}}
\affiliations{\textsuperscript{\rm 1}Computer Science, New York University Abu Dhabi, Abu Dhabi, UAE\\
\textsuperscript{$*$}Corresponding author: hazem.ibrahim@nyu.edu}

\begin{document}
\maketitle

\begin{abstract}
Studies of careers in science, film, music, and books report a common pattern. When a person's most successful work arrives is close to a random draw over the works they produce. How large their successes tend to be, in contrast, follows a stable, person-specific factor. We test whether this pattern holds for open-source software careers and whether it changed when generative AI coding tools arrived. From the complete public record of GitHub push events (2015--2025), we reconstruct 102.2M career works by 6.15M contributors, and for the 908k contributors whose repositories publish packages, we measure each work's impact by how many downstream packages come to depend on it. First, we find that the timing of a career's biggest hit is close to a lottery over their works, as in science and the arts, with a small, replicable lean toward early career that grows as careers get longer. Second, some coders reliably produce higher-impact work than others, but this lasting personal factor accounts for only part of why impact persists (about a fifth in our primary specification); the rest behaves like momentum, success feeding on itself for a period of time. Third, within the same contributors, this structure did not change after ChatGPT's release. The stable factor's weight grew by about as much as it grew for an earlier cohort that simply aged, and subtracting the effect of aging from the effect of generative AI puts the shift at +0.03 (95\% CI $[-0.22, +0.23]$), indistinguishable from zero. The success pattern documented in science and the arts therefore describes open-source careers too, and it shows no detectable break across the arrival of generative AI. These results have implications for how track records on open platforms should be read and on what to expect from generative AI for the careers built on them.
\end{abstract}

% ===========================================================================
\section{Introduction}
% ===========================================================================

Open-source track records increasingly function as public résumés. Employers read GitHub profiles as signals of ability \citep{marlow2013,goldbeck2025}, and the platform's transparency was designed to make such inferences easy \citep{dabbish2012}. Whether those inferences are sound depends on a question the platform cannot answer by inspection. How much of a visible open-source career reflects a stable property or ``quality'' of the person, and how much of it reflects chance and accumulated position? Literature in the science of success gives this question a quantitative form \citep{fortunato2018,clauset2017}. Merton's Matthew effect, the observation that success flows disproportionately to the already successful \citep{merton1968}, and its formalization as cumulative advantage \citep{price1976,petersen2011} explain why success concentrates, but they do not answer how much the individual themselves contributes to their long-term success.

Two findings from the literature turn this into a measurable question. In science, a researcher's highest-impact paper is equally likely to be any paper in their career, a finding known as the random-impact rule. How large each success is, in turn, splits into two parts: work-level luck times a stable person-specific factor, a decomposition formalized in the $Q$-model of \citet{sinatra2016}. Together these findings describe a success pattern in which chance governs \emph{when} a person's biggest hit arrives and a persistent individual factor predicts \emph{how big} that success can be. The same pattern has been documented for film, music, and books \citep{janosov2020}, along with the related hot-streak phenomenon, in which a person's most successful works cluster together at random points in the career \citep{liu2018,liu2021}. Experimental and modeling work have further supported this division between luck and skill, showing that the size of a hit is unpredictable even when quality is held fixed \citep{salganik2006,simkin2020}.

The domains where this pattern has been documented, however, are all built the same way. In science, film, music and books alike, a person's output is a series of discrete works with a fixed list of named authors, judged by an audience or by citations. Open-source software differs considerably. A repository accumulates commits from many contributors over its lifetime, so credit has no fixed boundary; impact spreads through dependencies between software packages \citep{zimmermann2019,decan2019} rather than through an audience; and the platform itself shapes what observers see. Judgments on the platform are also known to depend on the author's visible identity and geography \citep{terrell2017,rastogi2018,wachs2022}, which is a reason not to read any stable person-specific factor as ability or quality. If the pattern of success survives in this very different setting, it is more general than the domains it was discovered in; if it breaks, code is the first documented creative domain where it fails. Here, we study contributors who publish packages, the segment of open source where impact can be measured through dependencies. This is a distinctive, infrastructural population, and our claims throughout this study apply to it.

Software also offers something no prior domain has. Midway through our observation window, a new technology was introduced, which altered the act of writing code itself, as generative AI assistants took off after GitHub Copilot's general availability (June 2022) and ChatGPT's release (November 2022). Controlled and field experiments find that these tools speed up individual tasks, with the largest gains often going to less-experienced workers \citep{peng2023,noy2023,brynjolfsson2025,cui2026,hoffmann2024}, although these gains are limited to the tasks the technology already handles well \citep{dellacqua2026}. Software development ranks among the occupations most exposed to these tools \citep{eloundou2024}. From this literature, two opposing predictions about careers can be made. First, if AI compresses skill differences, individual advantage should erode (the democratization view). On the other hand, if AI multiplies the output of those who already have the position and skills to put it to work, advantage should concentrate (the rich-get-richer view). Both predictions concern how success is divided across people, not how much of it there is in total. Existing studies measure the second quantity; a tool can make everyone faster while leaving unanswered whether the same people still end up on top and how much of their success still comes down to chance.

Therefore, in this study, we ask three questions:
\begin{itemize}\itemsep2pt
\item \textbf{RQ1.} Does the random-impact rule hold for open-source package careers, i.e., is the timing of a contributor's biggest hit a lottery over their works?
\item \textbf{RQ2.} How much of the persistence in a contributor's impact is a stable individual factor, as opposed to cumulative advantage?
\item \textbf{RQ3.} Within the same people, does the luck--skill structure of success look different in the generative AI era than before it?
\end{itemize}

Throughout, we interpret the stable factor as a \emph{positional-advantage factor} $\Phi$, a fixed person-specific level (in the $Q$-model sense) that folds together skill, accumulated visibility, and platform position. This ambiguity is not specific to our setting; \citet{sinatra2016} note that their model is unchanged if $Q$ reflects factors such as education or institution rather than ability, and on GitHub, where judgments demonstrably depend on the author's visible identity and geography, the positional reading is more conservative.

A study with this many free analytic choices, such as which impact measure, which career-length floor, or which credit allocation to use, can manufacture any answer \citep{steegen2016,simonsohn2020}. Therefore, we pre-registered the full analysis plan,\footnote{The pre-analysis plan is registered on OSF at \url{https://osf.io/ruc7h/}.} enumerating every choice in advance into a grid of specifications, all of which are reported with a standard correction for testing many hypotheses at once (Benjamini--Hochberg) \citep{benjamini1995}. Contributors were split 50/50 into a discovery half used to explore and a held-out half reserved to confirm the hypotheses \citep{nosek2018}. Appendix~\ref{app:prereg} reports all pre-registration details, including the public registration documents (\url{https://osf.io/ruc7h/}), their amendments, and the handful of analyses added after the plan was submitted (marked with daggers in the tables).

\textbf{Contributions.} (1) The first test of the random-impact rule and the $Q$-model decomposition in software careers runs on open-source package contributors at a scale of $\sim$14,000 estimable careers per split half, with the main patterns replicated in the held-out half. (2) A decomposition of package-career impact into luck, a stable personal factor and momentum, built for the features that make software careers hard to measure: most works have zero impact, credit is shared across contributors, and calendar time distorts comparisons. We also show where the early lean and the stable share are largest and that the findings survive a full replacement of the measurement pipeline. (3) The first within-person, placebo-controlled comparison of the structure of success before and during the generative-AI era in a creative labor market, including a direct test of the era contrast and controls for career aging and for recent works that have not had time to gather impact.

% ===========================================================================
\section{Related Work}
% ===========================================================================

\subsection{Success in creative careers}
Our first two research questions ask whether a career's shape reflects the person or their luck, and careers in science and the arts already have a quantitative answer. Cumulative-advantage models explain why success concentrates. Early luck compounds into a rich-get-richer process \citep{merton1968,price1976}, visible in career longevity \citep{petersen2011} and in funding decisions, where applicants who narrowly win a grant go on to accumulate more than twice the follow-on funding of narrow losers \citep{bol2018}. What these models leave open is how much the individual contributes. The random-impact rule and the $Q$-model \citep{sinatra2016} answer this question with a decomposition. The timing of a career's biggest hit is random, and the size of each success is work-level luck multiplied by a stable person-specific factor; this decomposition reproduces the impact distributions of $\sim$2,900 physics careers. The same structure holds for film, music, and books \citep{janosov2020}. Hot streaks, periods in which a person's best works arrive back-to-back, add a complementary pattern where such streaks occur at random career positions and are preceded by exploration \citep{liu2018,liu2021}. Software careers are thus far absent from this literature, and our first two research questions bring them in.

Yet, there are three reasons why the stable factor introduced in the $Q$-model should not be read as an individual's ``ability'' or ``quality''. First, \citet{sinatra2016} themselves note that the dependence of $Q$ on exogenous factors such as education and institution remains unknown, and that the model is mathematically unchanged if $Q$ reflects them. Second, experimental cultural markets show that social dynamics decouple success from quality \citep{salganik2006}, and third, a preprint critique argues that the $Q$-factor tracks promotion more than product quality \citep{simkin2020}, a reading close to our own positional-advantage interpretation.

\subsection{Open-source careers and platform evaluation}
Whether this success pattern transfers to software depends on how open-source careers are produced, judged, and measured, and three lines of work, active since early case studies of Apache and Mozilla \citep{mockus2002}, shape our design. The first treats GitHub as a public r\'esum\'e where activity on the platform is visible to anyone, allowing observers to read it as evidence of skill and commitment \citep{dabbish2012}. Employers evaluate GitHub profiles in hiring \citep{marlow2013}, and contributors ramp up their visible work when a job search approaches \citep{goldbeck2025}. A second line shows that judgments on the platform are not about the code alone. Pull requests are accepted at different rates depending on the author's inferred gender \citep{terrell2017} and country, with reviewers favoring authors from their own country \citep{rastogi2018}, and contributions tend to cluster in a few world regions even though the work could plausibly be done from anywhere \citep{wachs2022}. These findings are a second reason we read the stable factor as positional advantage rather than ability. A third line maps how impact travels between projects; most of the ecosystem depends on a small core of packages \citep{zimmermann2019,decan2019}, which is why we measure a work's impact by the downstream packages that come to depend on it, and why maintaining such a package is a career in itself \citep{eghbal2020}. None of this work measures how much of an individual career is luck and how much is a stable property of the person. That gap is what this study aims to fill.

\subsection{Generative AI and work}
If generative AI has changed anything about this pattern, it should enter through the stable factor. A tool that raises everyone's capacity to produce good work would compress the person-specific differences the $Q$-model measures. On the other hand, a tool that multiplies the output of those who already have the position and skills to put it to work would widen them. Evidence from the literature thus far leaves both scenarios open. A controlled experiment finds 56\% faster task completion with AI coding assistance \citep{peng2023}. Three enterprise field experiments find a 26\% increase in completed tasks, concentrated among less-experienced developers \citep{cui2026}. Under Copilot's staggered rollout, open-source maintainers reallocate effort toward coding and away from project management \citep{hoffmann2024}. Beyond code, generative AI compresses worker skill gaps in writing tasks \citep{noy2023} and customer support \citep{brynjolfsson2025}, with gains limited to the tasks the technology already handles well \citep{dellacqua2026}. Software development ranks among the most exposed occupations \citep{eloundou2024}. All of these observe tasks over weeks or months. None observe careers, and none ask whether the division of success across people changed. Our third question fills that gap by comparing the same people before and after a fixed calendar date and testing how success is divided.

% ===========================================================================
\section{Data and Measurement}
% ===========================================================================

\subsection{Data sources and coverage}
Our career data come from GitHub Archive \citep{gharchive}, a public log of everything that happens on GitHub, from which we extract every pushed commit from January 2015 through September 2025. While additional data beyond this end date would have been valuable for our analyses, GitHub removed commit-level author information from the public stream on October 8, 2025, and after that date, research data at the commit level cannot be rebuilt from any public source.

Information about which packages depend on which comes from two open datasets, the Libraries.io snapshots (five releases, June 2017 to January 2020) \citep{librariesio} and the Ecosyste.ms dumps (eight releases, August 2022 onward) \citep{ecosystems}. Neither source covers January 2020 to August 2022. We bridge that gap by interpolating between the snapshots on either side (in log space, with every bridged row flagged), and we test whether the bridge has an impact on our results with two sensitivity analyses (Results; Appendix~\ref{app:sensitivity}).

\subsection{Github identity resolution}
GitHub Archive hides the personal part of each commit author's email address behind a hash, so we cannot simply match addresses. Instead we build a dictionary: for every platform account we compute the hashes its commits could carry (GitHub's two no-reply address formats, in both original and lowercased form, built from the account's login and numeric id), and match commit hashes against it, so a commit author is decoded exactly or not at all. When this does not produce a match, we assign an author to the account that pushed at least 80\% of that author's commits within the era ($\geq$5 commits over $\geq$2 months), and otherwise leave the commit unassigned. We then apply a bot filter to remove automation accounts, so that machine pushes are never credited as human work. We validate our results against a gold set of 28,896 account--identity pairs whose true owner is known from ORCID records. Precision, the share of our assignments that are correct, is 1.000 for 2015--19 commits, with zero misattributions in any world region, and stays $\geq$0.98 in every era and region after the bot filter; Appendix~\ref{app:identity} details the recent-era complications (no-reply addresses, continuous-integration pushes) that the dictionary and the filter handle. We resolve 60.6\%, 71.8\% and 74.7\% of commits in 2015--19, 2020--22 and 2023--25, and leave the rest unassigned, such that identity errors cost us coverage rather than correctness.

\subsection{Measuring works, credit and impact}
We define a \emph{work} as one person's contribution to one repository, dated by their first commit to it ($\tau$), and define a career as a person's works in chronological order. Most repositories have several contributors, so a work's impact has to be divided among them. We do this under three different credit rules, following the credit-sharing checks of \citet{sinatra2016}. \textit{R-share}, our primary rule, splits credit in proportion to each person's share of the commits. \textit{R-lead} gives the work to its largest committer alone, and \textit{R-core} splits it equally among the committers who together wrote 80\% of it. For robustness, we never headline a result whose direction flips across the three rules, and none of the results documented below do.

We define the \emph{impact} of a work as how many packages come to depend on its repository's packages, counting packages that depend on them directly or through one intermediary (reach$^2$; the direct count alone, in-degree, is the alternate measure). Impact is evaluated $W$ months after the work began, $W\in\{24,36,60\}$. We note that raw dependency counts are not comparable across communities or years. An npm package, for instance, collects dependents on a different scale than a Cargo or RubyGems package, and later entrants join denser ecosystems. We therefore log-scale each work's impact and center it on the median of works published in the same year and in the same ecosystem, allowing us to compare a work's score relative to its contemporaries.

\begin{figure*}[t]
\centering
\includegraphics[width=\textwidth]{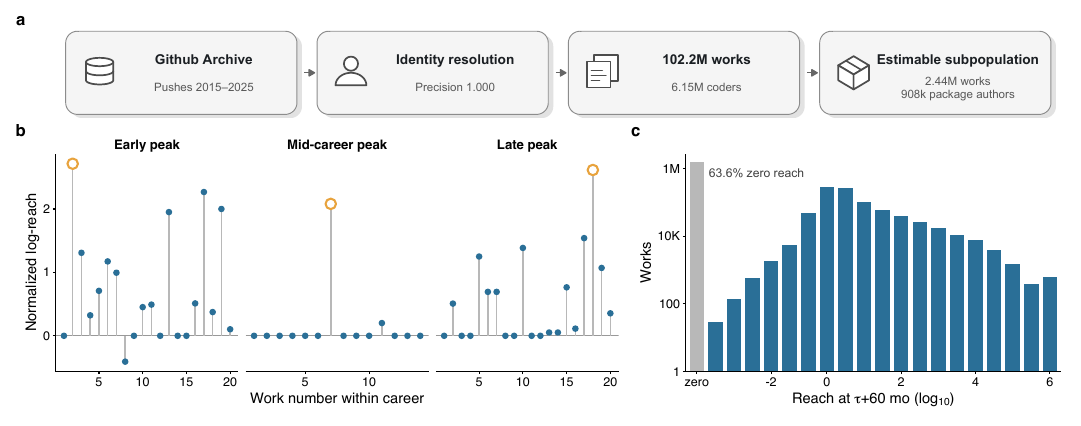}
\caption{\textbf{Measuring careers in code.} \textbf{a}, The pipeline. GitHub Archive push events (2015--2025, frozen September 30, 2025) are matched to platform accounts (the hash dictionary and 80\% backup rule of Methods; precision 1.000 on the 2015--19 gold set, $\geq$0.98 in every later era and region, Table~\ref{tab:identity}), yielding 102.2M works by 6.15M contributors; the 2.44M works (908k contributors) whose repositories publish packages are the population we can estimate on (the ``estimable subpopulation''). \textbf{b}, Three real (anonymized) careers where each vertical line is one work, its height the work's impact at $\tau+60$ months (log-scaled and centered on the same-year, same-ecosystem median); the golden ring marks the career's biggest hit. \textbf{c}, Distribution of impact at $W$=60: the grey bin at left holds the 63.6\% of works with zero reach (63.0\% exactly zero; the bin also holds works whose credit-shared impact is near zero).}
\label{fig:construct}
\end{figure*}

This process yields 102.2M works by 6.15M contributors (Fig.~\ref{fig:construct}a). Impact through dependencies is only measurable where a repository publishes packages, so the population we can estimate on is 2.44M works by 908,417 contributors across 1.82M repositories. Most works never gain measurable impact; 63.0\% reach zero downstream projects even at the widest horizon, $W$=60 (56\% within the primary sample; Fig.~\ref{fig:construct}c). We keep the zeros in the analysis rather than discard them, at the cost of underestimating any stable factor. Our main analysis condition holds 14,091 careers in the discovery half, roughly five times the sample of the original random-impact study on careers in science \citep{sinatra2016}.

\subsection{Experimental Design}
Crossing the three credit rules (R-share, R-lead, R-core), the three horizon lengths ($W\in\{24,36,60\}$), three minimum career lengths ($\geq$3/5/10 works) and the two impact measures (reach$^2$, in-degree) gives us 54 different conditions. We run and report all 54 conditions after correcting for multiple testing with the Benjamini--Hochberg procedure, and designate one condition as primary (R-share, $W$=60, $\geq$5 works, reach$^2$), fixed in the pre-registration before any data were analyzed. Each component of the primary condition is the natural default of its dimension: R-share is the closest analogue of the credit sharing in \citet{sinatra2016} and sits between the winner-take-all and equal-split alternatives; the 60-month horizon gives impact the most time to accumulate while remaining fully observable; the $\geq$5-work floor is the middle of the three; and reach$^2$ captures spread through the dependency graph beyond a package's immediate dependents. Contributors were split in half before any analysis. We explored on one half (the discovery half), then ran the identical analysis once on the other (the held-out half). Every confirmatory number in this study comes from that held-out half.

% ===========================================================================
\section{Methods}
% ===========================================================================

\subsection{RQ1: the lottery test}
For each career, $P_i$ is the position of the person's highest-impact work (the ringed works in Fig.~\ref{fig:construct}b) divided by their career length. If timing is a lottery, the biggest hit is equally likely to land anywhere; thus, for a career of $N$ works its position is uniform on $\{1,\dots,N\}$, and pooling this prediction over the career lengths present in a condition gives the distribution we test against. Ties on impact are broken at random, and careers whose works all tie (all-zero reach) are excluded, a choice we validated on synthetic careers (Appendix~\ref{app:validation}). A departure from the lottery means the biggest hits land systematically earlier (or later) in careers than a random draw would put them. We summarize this departure in three ways: (1) the mean of $P$, which falls below 0.5 when hits come early and rises above 0.5 when they come late; (2) the total variation distance (TVD) between the observed and predicted decile distributions, that is, the share of careers that would have to change decile for the observed distribution to match the prediction; and (3) a permutation $p$ value from 5,000 within-career shuffles of the hit position. A $\chi^2$ test against the exact expected decile counts is computed alongside as a consistency check.

\subsection{RQ2: stable-factor decomposition}
If two works by the same person tend to have similar impact, that similarity can come from two very different sources. The person may simply be the kind of contributor whose work reliably lands well, a property that transfers from project to project and never wears off. Alternatively, one success may temporarily lift the next, through visibility, adoption, or placement in a growing ecosystem, and that boost fades as the works grow farther apart (i.e., momentum). The two sources leave different fingerprints in how impact correlates across a career, which the following fit separates. For each condition we compute $r(k)$, the correlation between the impacts of a coder's works $k$ positions apart (on a normalized log scale), pooled across contributors, and fit
\begin{equation}
r(k) \;=\; \stableshare \;+\; B\,\rho^{k}
\end{equation}
by constrained least squares ($\stableshare,B\geq0$; $0\leq\rho\leq0.9$). A permanent property keeps even distant works correlated, so it shows up as the flat plateau $\stableshare$, the share of impact variance carried by a stable coder effect ($\Phi$). Momentum, on the other hand, dies out as works get farther apart, so it shows up as the decaying term $B\rho^{k}$. The fit caps $\rho$ at 0.9 because a decay slow enough never to fade would be indistinguishable from a permanent effect; this cap is never reached in our analysis (maximum fitted $\rho$=0.68 across the 54 conditions; Appendix~\ref{app:prereg}). Confidence intervals come from a coder-level bootstrap (1,000 resamples).

A fitted plateau alone is not enough to claim a stable factor, because the fit can return a positive $\stableshare$ for the wrong reasons: plain noise, a small plateau sitting under a much larger momentum term, or a career whose spread keeps growing. We therefore count a stable factor as genuine only if it passes three pre-registered checks, one for each of these traps. First, the plateau must be large enough to matter ($\stableshare>0.10$, with bootstrap lower bound above 0.02), so that noise alone cannot produce it. Second, the plateau must account for at least a fifth of the total work-to-work correlation ($\stableshare/(\stableshare+B)>0.20$; we call this the \emph{persistence ratio}), so that we never call a career stably different when momentum explains almost all of it. Third, the spread of impact within a career must stay roughly constant, with the variance of late works no more than 1.5 times that of early ones, so that a career that simply keeps growing cannot be mistaken for a permanent property of the person. Tables and the appendix abbreviate the RQ1 lottery test as C1 and this decomposition as C2.

We validated the decomposition in three ways. First, before analyzing any real career data, we tested the estimators on nine families of synthetic careers, confirming that they recover a stable factor when one is present and report none when it is absent (Appendix~\ref{app:validation}). Second, a split-half check divides each career into two random halves and correlates the coder's average impact across them, testing for a stable person effect without using the order of works at all. Third, we estimate per-coder effects separately under the two impact measures (reach$^2$ and in-degree) and correlate them; if both measures read the same underlying factor, they should rank people the same way, and we claim a single stable factor only if that rank correlation clears $\rho\geq$0.7 (Results; Appendix~\ref{app:prepost}).

\subsection{RQ3: era-comparison design}
\emph{Incumbents} are contributors with qualifying careers on both sides of a divide at January 2023, that is, in the pre-AI era (January 2015 to December 2019) and in the post-AI era (January 2023 to September 2025; below, the pre-era and post-era), where a qualifying career has at least the condition's minimum number of works ($\geq$5 at the primary condition) spanning at least 24 months. We do not directly observe who adopted AI tools, and adoption is not observable at our scale: the public record only carries the metadata of each push, not the code itself, and even with code in hand there is no reliable way to tell AI-assisted work from unassisted work. Our comparisons are therefore between calendar eras within the same people, so the conclusions concern the generative-AI era, not AI adoption itself. To compare the eras, we run the identical estimators separately on each side of the divide, with the horizon shortened to $W$=24 or 36 months on both sides, so that post-era works, which have had less time to gather impact, are never measured over a shorter window than the pre-era works they are compared with. Crossing the remaining dimensions of the grid gives 36 pre/post conditions; the primary one is R-share, $W$=24, $\geq$5 works, reach$^2$. Each incumbent stays in the same discovery or held-out half as in the main analysis.

A raw before/after comparison has a built-in problem: every incumbent is necessarily older after any divide, so a change across it could reflect career aging rather than the era itself. We therefore construct a \emph{placebo divide} of identical design on people the AI era cannot have touched: contributors who entered in 2015--16, split at July 2017, with a post-window of the same length. The placebo shows the effect of aging alone, and real-divide changes are compared against it rather than against zero.

The end of the data poses a second problem. A work made shortly before the data freeze has not had time to gather impact, so we drop any works whose measurement date $\tau+W$ falls after the September 30, 2025 freeze. Post-era estimates therefore rest only on fully measured works ($\approx$49--52\% of post-era works are dropped at $W$=24; the share is reported per condition), while the placebo eras are fully observed. Whether this imbalance matters is itself tested by a matched-censoring placebo, which imposes the identical rule on the placebo cohort at a pseudo-freeze of April 2020, cutting its post-window at the same point the real freeze (September 2025) ends our dataset.

In addition to the descriptive real-vs-placebo comparison, we test the era contrast directly. We define $D$ as $D=(\Delta\stableshare)_{\text{real}}-(\Delta\stableshare)_{\text{placebo}}$, or the real divide's pre-to-post change in the stable share minus the placebo's. If the AI era changed nothing beyond aging, $D$ should be near zero. We estimate $D$ with a contributor-level bootstrap that resamples each person with their pre and post sequences kept together (1,000 resamples at the primary condition), and evaluate it with two tests. A confidence interval asks whether $D$ differs from zero, and an equivalence test (TOST) asks whether $D$ is provably small. This equivalence test passes only if the 90\% CI lies entirely within $\pm$0.10, the smallest shift we count as meaningful. Since the AI era could also have shifted \emph{when} hits arrive, not just how large the stable share is, an analogous statistic $T$ applies the same real-minus-placebo contrast to the timing departure (TVD).

% ===========================================================================
\section{Results}
% ===========================================================================

\subsection{RQ1: Hit timing is close to a lottery}

\begin{figure}[t]
\centering
\includegraphics[width=\columnwidth]{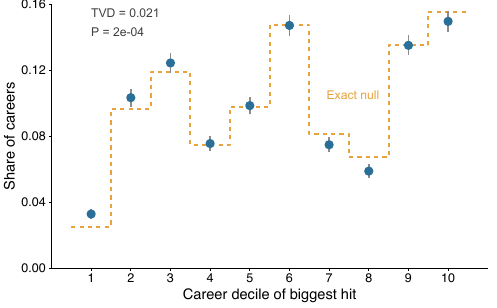}
\caption{\textbf{Primary-condition decile view of hit timing.} Share of careers whose biggest hit falls in each career decile (points; bars are 95\% CIs) against the lottery prediction (dashed amber step, labeled ``exact null'' in the panel), computed from the observed mix of career lengths; TVD 0.021, permutation $p=2\times10^{-4}$, $n$=12,270 careers entering the test (discovery half; held-out counterpart in Appendix Table~\ref{tab:ed1}). The prediction is jagged rather than flat because careers have whole-number lengths, so deciles collect unequal shares of the possible hit positions. The points sit slightly above the prediction in the early deciles and at or below it in the late ones, the early lean reported in the text.}
\label{fig:luckdecile}
\end{figure}

\begin{figure*}[t]
\centering
\includegraphics[width=\textwidth]{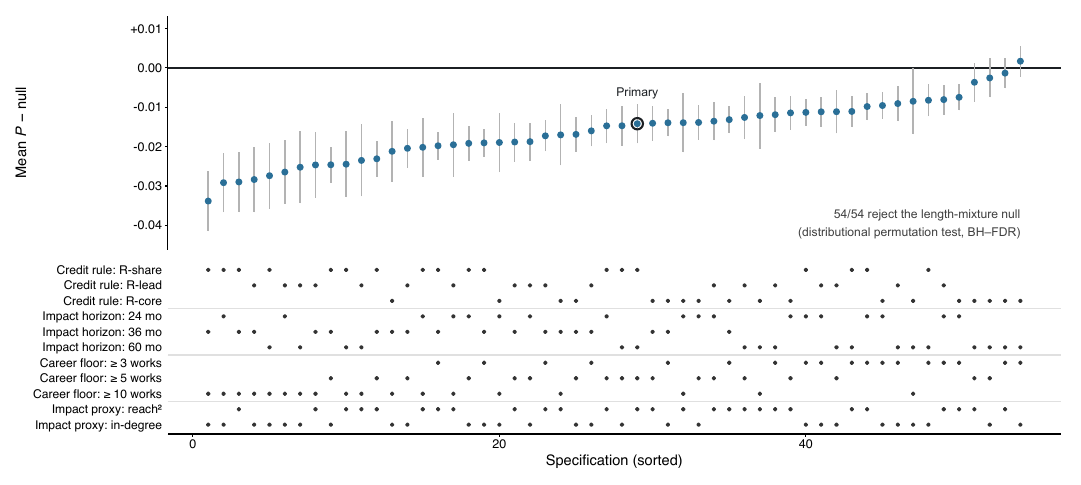}
\caption{\textbf{Hit timing is close to a lottery.} Departure of the mean biggest-hit position from its lottery expectation (``Mean $P$ $-$ null'' in the panel) across all 54 conditions, sorted; bars are 95\% CIs, and the dot-matrix below maps each condition (credit rule, horizon $W$, career floor, impact measure). All 54 conditions reject the lottery prediction (the ``length-mixture null'' of the panel) under the permutation test with the Benjamini--Hochberg correction, with uniformly small departures (median TVD 0.028 in the discovery half, 0.029 held-out). The test uses the full decile distribution, so conditions whose \emph{mean} departure sits near zero (right edge) still reject: early and late deviations cancel in the mean but not in the TVD. The ringed primary condition is shown decile-by-decile in Fig.~\ref{fig:luckdecile}.}
\label{fig:luck}
\end{figure*}

RQ1 asks whether a contributor's biggest hit is equally likely to arrive at any point in their career, as it is in science and the arts. We find that it very nearly is. At the primary condition, the observed positions of biggest hits sit close to the lottery prediction and lean slightly early; the mean position is 0.548 against a lottery expectation of 0.562 (above one half because careers are finite: a uniform draw over $N$ works has mean position $(N{+}1)/2N$). The departure is small but statistically unambiguous (TVD 0.021, permutation $p=2\times10^{-4}$; Fig.~\ref{fig:luckdecile}), and it is small everywhere. All 54 conditions reject the lottery prediction after correction for multiple testing, with median TVD 0.028 (max 0.052) and an early lean in 53 of 54 (Fig.~\ref{fig:luck}). The held-out half yields the same results, with a mean 0.550 vs.\ 0.562 at the primary condition (TVD 0.022, $p=2\times10^{-4}$), 54/54 rejections, and median TVD 0.029 (Appendix Table~\ref{tab:ed1}). In other words, when a contributor's biggest hit arrives is essentially a lottery, with a slight tilt toward the start of the career.

\textbf{Why do we see an early lean in hit timing?} Having established that the departure from the lottery is a lean toward early works, we next ask why. The lean grows mainly with career length (Table~\ref{tab:mod}). Raising the career-length floor from $\geq$3 to $\geq$5 to $\geq$10 works raises the median TVD from 0.019 to 0.023 to 0.041 and roughly doubles the median early shift, the gap between the observed mean hit position and its lottery expectation, from $-0.011$ to $-0.024$. In other words, the longer a career, the further its hit timing sits from a lottery. This is what an attention advantage for early works would produce, whether because repositories collect the most attention early in their life or because early works entered a less crowded ecosystem; the longer a career runs, the more late works accumulate that never overtake an early hit. The early lean appears under every credit rule and both impact measures (slightly larger under R-share and in-degree) and is largest at the middle horizon ($W$=36), an ordering the held-out half repeats (median shift $-0.014$, $-0.020$, $-0.010$ for $W$=24/36/60). Our sample size is large enough that even this small departure is statistically significant, but it remains small in practical terms: a TVD of 0.02--0.04 means that only two to four careers in a hundred sit in a different decile than the lottery predicts. Contributors, in other words, have essentially no control over when their biggest hit arrives, and what little timing structure exists is consistent with the extra attention early works accumulate rather than with anything contributors do differently across their careers.

\subsection{RQ2: A stable factor, but a minority share}

\begin{figure*}[t]
\centering
\includegraphics[width=\textwidth]{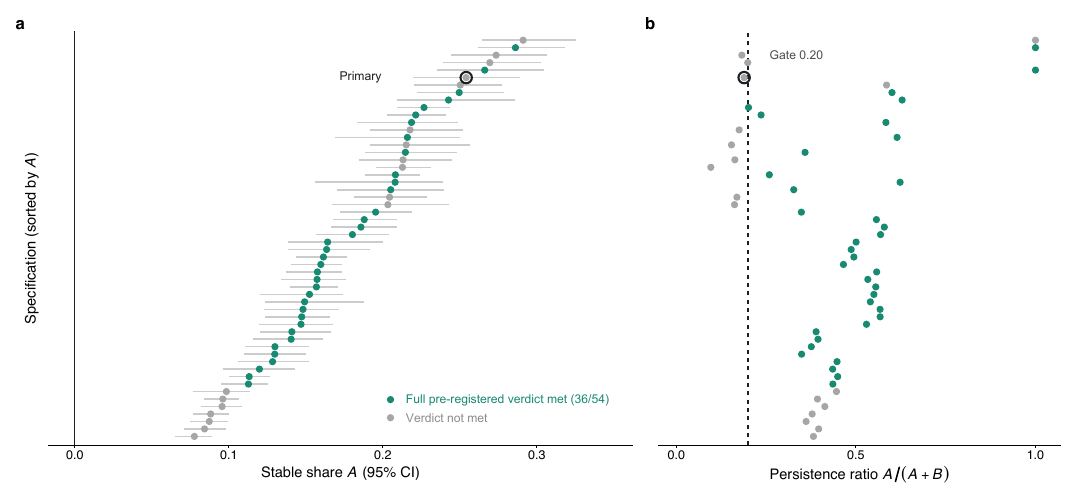}
\caption{\textbf{A stable coder effect exists everywhere but carries a minority of persistence.} \textbf{a}, Stable share $\stableshare$ with bootstrap 95\% CIs for all 54 conditions in the discovery half, sorted; the CI excludes zero in every condition. \textbf{b}, The persistence ratio $A/(A{+}B)$ against the 0.20 threshold (the dotted ``Gate 0.20'' line), same row order. Teal marks the 36 of 54 conditions meeting the full three-criterion verdict (33 of 54 in the held-out half). The ringed primary condition sits at ratio 0.189 (0.180 held-out), marginally below the threshold; the calendar sensitivity analyses of the text move it above the threshold in both halves. The primary-condition decomposition and split-half consistency are in Appendix Fig.~\ref{fig:skilldecomp}.}
\label{fig:skill}
\end{figure*}

RQ2 asks how much of the persistence in a contributor's impact reflects a stable property of the person, informally the intrinsic ``quality'' of a coder, as opposed to momentum that fades. We find that a lasting personal factor exists in every one of the 54 conditions, but it explains only part of why impact persists: about a fifth at the primary condition, with fading momentum accounting for the rest. Concretely, $\stableshare$, the share of impact variance the stable factor explains, is above zero in all 54 discovery conditions (median 0.172) and all 54 held-out conditions (median 0.201). At the primary condition the two halves agree almost exactly ($\stableshare$=0.254 [0.220, 0.289] discovery; 0.251 [0.209, 0.292] held-out), and the persistence ratio $A/(A{+}B)$ is 0.189. In other words, of everything that makes one work's impact predict the next, roughly 19\% comes from the stable factor and the rest from momentum. The full three-criterion verdict outlined in the Methods section is a stricter test and it passes in 36 of 54 discovery conditions and 33 of 54 held-out ones (Fig.~\ref{fig:skill}). The primary condition itself falls marginally short in both halves: its persistence ratio sits just under the 0.20 threshold (0.189 discovery, 0.180 held-out), and in the held-out half the spread of impact also grows too much over the career (growth ratio 1.75 against the 1.5 limit; 1.42 in discovery).

\textbf{Why the primary condition fails.} The primary condition fails its verdict because too much of the work-to-work correlation reads as momentum: the persistence ratio lands at 0.189, just under the 0.20 threshold. Part of that momentum appears to be an artifact of the data rather than of careers. No dependency snapshots exist between January 2020 and August 2022, so every impact value measured across those years is interpolated (Data and Measurement), and interpolated values are smoother than real ones, which is what momentum looks like to the fit. We therefore re-ran the decomposition without the interpolated stretch, using two methods: (1) keeping only careers whose measurement windows sit entirely within snapshot-covered years, and (2) keeping all careers but dropping the individual measurements that cross the gap. The two approaches agree. In the discovery half the persistence ratio rises to 1.00 and momentum vanishes entirely; in the held-out half it rises to 0.346, clearing the threshold, though the verdict there still fails, now at the variance-growth limit (1.53 vs.\ 1.5). The lottery result does not move in any re-run. In short, the interpolated years inflate momentum at the stable factor's expense. Appendix~\ref{app:sensitivity} reports both analyses in full.

\textbf{How measurement choices move the stable share.} Across the conditions (Table~\ref{tab:mod}), the stable share we estimate is larger when impact is measured deeper in the dependency graph (median $\stableshare$ 0.215 for reach$^2$ vs.\ 0.157 for in-degree; held-out 0.240 vs.\ 0.187) and is largest under proportional credit allocation (R-share 0.235 $>$ R-lead 0.182 $>$ R-core 0.117). Both gradients fit the positional-advantage interpretation. The persistent part of a contributor's impact depends on how far their work spreads through the dependency graph and on how many projects they share credit in, and these are exactly the quantities platform position governs. The two impact measures nonetheless agree on who the consistently high-impact coders are. Per-person effects under reach$^2$ and under in-degree correlate at Spearman $\rho$=0.979 in both halves, so the two measures read a single person-level factor on different scales. Reliability, by contrast, depends on career length. Split-half consistency, the agreement between a coder's average impact in one random half of their career and the other, rises from $r$=0.29 at the $\geq$3-work floor to $r$=0.53 at $\geq$10 while median $\stableshare$ barely moves (Appendix Fig.~\ref{fig:skilldecomp}). Longer careers, in other words, measure the stable factor more reliably.

Our results leave no doubt that some coders reliably produce higher-impact work than others. What is more uncertain is \emph{why} a coder's impact persists, whether it reflects a permanent property of the person or success feeding on itself for a while. That division swings with which half of the data and which calendar handling we use (the stable factor's share of persistence reads 0.18, 0.35 or 1.00 across those choices), so we treat it as less settled. The estimates themselves come with two systematic pulls in opposite directions. Works with zero impact drag the estimate down, so the true stable share is, if anything, larger than we report. Moreover, because the stable factor $\Phi$ folds in platform position as well as skill, the part of it that reflects true skill is, if anything, smaller. The summary that survives every analysis is luck with a stable floor. A reproducible individual factor exists but explains a minority of the variance in hit size; the remainder is momentum and work-level chance.

\subsection{RQ3: No detectable break in the generative-AI era}

\begin{figure*}[t]
\centering
\includegraphics[width=\textwidth]{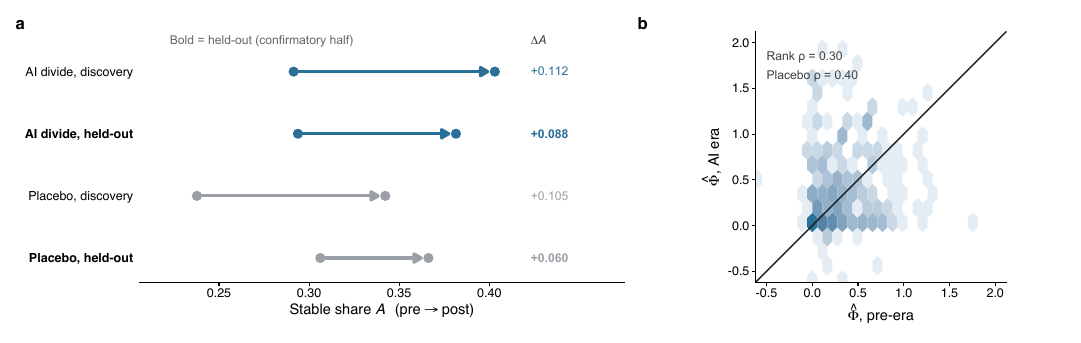}
\caption{\textbf{No detectable break in the generative-AI era.} \textbf{a}, Stable share $A$ before and after the January 2023 divide for the same incumbents (blue), next to the placebo aging divide (grey). Bold marks the held-out (confirmatory) half; Appendix Table~\ref{tab:ed2} lists the per-era 95\% CIs. The real cohort's rise (+0.088) is about the size of the rise aging alone produces (+0.060), the direct contrast $D$=+0.035 $[-0.218, +0.230]$ covers zero, and the rise disappears once censoring is equalized or careers are compared at the same position (see text). \textbf{b}, Estimated per-coder effects $\hat{\Phi}$ before vs.\ after the divide (hex-binned; held-out incumbents with $\hat{\Phi}$ estimable in both eras, $n$=417 of 421). The ordering of coders carries across the divide at rank $\rho$=0.30 (placebo 0.40). Hit timing stays lottery-like in both eras (Appendix Fig.~\ref{fig:ailottery}).}
\label{fig:ai}
\end{figure*}

Our final question asks whether this luck--skill structure looks different in the generative-AI era. We compare the same people before and after the January 2023 divide and detect no change beyond what career aging produces on its own. For the stable share, in the held-out half at the primary pre/post condition (421 incumbents), the estimate rises from 0.294 [0.206, 0.368] to 0.381 [0.243, 0.464] across the divide, a change of +0.088. On its own, that rise could be interpreted as an effect of ChatGPT's release. However, aging alone produces a rise of similar size: the placebo cohort, which crossed its own divide years before AI arrived, rises by +0.060. The same holds across all 36 pre/post conditions, where the real cohort's median rise of +0.039 sits inside the range of rises the placebo cohort produces (Appendix Table~\ref{tab:ed2}, Fig.~\ref{fig:ai}a). Subtracting the placebo cohort's rise from the real cohort's, condition by condition (the statistic $D$ described in Methods), leaves $D$=+0.035 at the primary condition, with a 95\% CI of $[-0.218, +0.230]$ that covers zero, as do all 72 $D$ CIs across both halves. These results therefore rule out era shifts in the stable share larger than about $\pm$0.23, but they are too wide to certify that smaller shifts are absent. Put plainly, the stable factor's weight grew after ChatGPT's release by about as much as it would have grown anyway, and any true era effect, if one exists, is too small for this design to detect.

For hit timing, biggest hits stay lottery-like in both eras; no real-divide condition rejects its own era's lottery prediction after correction in either half (held-out primary TVD 0.05 pre and 0.07 post, $p>0.5$; Appendix Fig.~\ref{fig:ailottery}). The real-minus-placebo contrast for timing further verifies this result ($T$=+0.005 $[-0.080, +0.117]$, covering zero). Lastly, the ordering of per-coder effects carries across the real divide about as well as it carries across the placebo divide (rank $\rho$=0.30 vs.\ 0.40 held-out; Fig.~\ref{fig:ai}b). In short, these results suggest that the same people stayed on top, their biggest hits stayed randomly timed, and their advantage grew no faster than aging alone predicts.

\textbf{Why the raw rise carries no information.} Two features of the design, both unrelated to AI, account for the +0.088 rise we estimate in full, and removing either one erases it (complete analyses in Appendix~\ref{app:prepost}). The first is censoring, the loss of works too recent to measure. Works made shortly before the September 2025 data freeze have had no time to gather impact, so about half of the real cohort's post-era works are dropped, while the placebo eras are fully observed; measuring careers on only their earliest post-era works inflates the fitted stable share by itself. Imposing the identical handicap on the placebo cohort raises its apparent gain to +0.176, overshooting the real rise, and turns the real-vs-placebo contrast slightly negative ($D$=$-0.089$ $[-0.540, +0.340]$). The second is career stage. Comparing each incumbent's post-era works with the same number of their own late pre-era works, so that both sequences sit at the same career position, removes the rise entirely ($\Delta\stableshare$=+0.000 $[-0.247, +0.271]$). Time would relax the first constraint on its own, since post-era works only need more years to accumulate dependencies, and a re-run of this comparison at the full 60-month horizon would be the natural follow-up; that re-run is possible, however, only if GitHub restores the public commit-level data access it removed in October 2025.

% ===========================================================================
\section{Discussion}
% ===========================================================================

Careers in science, film, music and books obey a common pattern. When a person's biggest hit arrives is essentially random, but how big their successes tend to be follows the person. We asked whether careers in open-source software obey the same pattern, and the short answer is yes, almost. In every condition, the timing of a contributor's biggest hit sits close to the lottery prediction, yet the data are large enough to show the fit is not perfect; hits lean slightly early, and the lean concentrates in long careers. That the pattern holds here is interesting, because software careers differ from the earlier domains in most of the ways that could plausibly break it. The work is collaborative, a platform mediates it, and success is judged through adoption rather than by an audience or by citations. A pattern that survives all of those changes looks like a general property of creative careers rather than an artifact of how citations behave \citep{wang2013}. The small early lean may be the more interesting finding. It is the one place where package careers depart from the earlier domains, and it gives future work a concrete mechanism to test, the extra attention early works collect through repository lifecycles and ecosystem entry timing.

GitHub profiles are read as public r\'esum\'es \citep{dabbish2012,marlow2013}, and our decomposition puts numbers on what a profile can actually tell an observer. A real, reproducible individual signal exists, but it explains only about a fifth of why impact persists from one work to the next; the rest is momentum that fades. The signal is also strongest exactly where platform position matters most, in how far work spreads through the dependency graph and how many projects share the credit, and reading it reliably takes a long career, since the split-half consistency of a career roughly doubles from short careers to long ones even though the stable share itself does not change. Meanwhile the single most visible event in a career, the biggest hit, is timed almost entirely by chance, and small early advantages compound as evaluators reward past success \citep{merton1968,bol2018}. The practical advice for anyone reading a profile \citep{goldbeck2025} is to look at consistency across many works, not the size or recency of one hit; inferring ability from hit timing is close to reading a lottery ticket.

Generative AI changes how much code gets written and how \citep{peng2023,cui2026,hoffmann2024}. We asked whether it changed who wins. Two familiar arguments suggest that it should, in opposite directions. On the democratization argument, AI compresses skill differences and erodes individual advantage; on the concentration argument, it multiplies the output of those already ahead. Within the same incumbent contributors, we see neither. The stable share rises after the divide by about as much as it rises for an earlier cohort that simply aged; the rise vanishes once the placebo cohort is handicapped by the same loss of too-recent works, and again once people are compared with themselves at the same career stage; and the direct contrast is centered on zero (+0.035), ruling out era shifts larger than about $\pm$0.23. However, we note that our design cannot certify that shifts smaller than $\pm$0.10 are absent, and it compares calendar eras, not measured AI adoption, on a modest sample (421 held-out incumbents at the primary condition). Within those limits, our interpretation is that generative AI can raise everyone's output while leaving intact who tends to end up on top and how much of that is chance.

\paragraph{Limitations.} Four constraints, each set by what public data make observable rather than by the design, limit our claims, and each points to concrete future work. The first is the population. Impact through dependencies is only measurable where repositories publish packages, so our claims cover package contributors, an infrastructural and visible segment holding about 2\% of reconstructed works, and dependency reach misses impact carried by applications, documentation, tooling, issue work and testing. Extending the decomposition beyond this segment will need impact measures for those forms of work at comparable scale. The second is the measurement calendar. Most works never gain downstream dependents, impact is observed only at snapshot dates, and the 2020--2022 lapse between the Libraries.io and Ecosyste.ms datasets had to be bridged by interpolation. The first two of those features push the stable-share estimates down, so the true stable share is, if anything, larger than we report; the interpolation bridge contributed to the one failed verdict, and the calendar sensitivities that contextualize it replicate in direction but not magnitude on the held-out half. Dense snapshot coverage has resumed since August 2022, so bridge-free re-estimation becomes possible as those calendars lengthen. The third is the era design. ChatGPT's release sits barely three years before GitHub's removal of public commit authorship closed the panel. ``The AI era'' is therefore a calendar period rather than measured adoption, the comparison is restricted to incumbents who remained active into the post-era, and the short, right-censored post-era limits how small an era effect the comparison can detect. These constraints should ease with time. Impact horizons keep maturing in later dependency snapshots even though the commit stream is closed, which will let the comparison detect smaller effects, and re-running the same design on post-2022 entrants as their careers complete would test democratization among the newcomers an incumbent design cannot see. The fourth is scope of inference. Grid-moderator readings are descriptive rather than confirmatory, and geographic representation in the resolved population is uneven, so we report no region-level analysis; the only regional numbers here validate identity-resolution precision. Two questions follow as future work on the panel built here: whether equal underlying work is credited equally across visible identity and inferred world region, which requires its own region-stratified validation, and what the positional advantage lines up with before a career begins.

\bibliography{references}

% Appendix tables/figures use A-prefixed numbering.
\setcounter{table}{0}
\renewcommand{\thetable}{A\arabic{table}}
\setcounter{figure}{0}
\renewcommand{\thefigure}{A\arabic{figure}}

\section{Appendix}

\subsection{Ethics Statement}
This study is observational secondary analysis of public platform data; no intervention was performed and no private data accessed. Under our institution's policy it is not human-subjects research and required no IRB review. Author emails in the source are hashed; our identity resolution links only public platform identities and is validated for precision so that abstention, not misattribution, absorbs uncertainty. All reported outputs are aggregate; example careers are anonymized; no individual-level identity, ranking or fault tables are released, and we release no classifier that infers personal attributes. Platform terms (research use of public, non-personal information with open-access publication) and data licenses and terms (CC-BY-SA for Libraries.io; open data for Ecosyste.ms; GH Archive is publicly archived GitHub event data used under GitHub's API terms) are respected. Geographic representation is uneven; we report no region-level results (regional breakdowns appear only in the identity-resolution validation), and we caution explicitly against ability readings of the stable factor. The paper's central interpretive stance is that measured ``coder quality'' is substantially position and momentum. The pre-registration, amendments, unseal attestation and a code-deviation note are public on OSF (\url{https://osf.io/ruc7h/}). Analysis code and aggregate result tables are available at \url{https://github.com/hazemibrahim97/coders-q} and will be archived on publication.

\subsection{Appendix Overview}
\noindent The appendix reports the full pre-registered specification grid in both halves (no condition is omitted), the full pre/post grid, the calendar-gap sensitivity in full, and the validation and provenance detail summarized in the main text. All tables are generated programmatically from the sealed result files; no number is transcribed by hand. Analysis code and the aggregate result files behind every figure and table are available at \url{https://github.com/hazemibrahim97/coders-q}.

\begin{figure*}[t]
\centering
\includegraphics[width=\textwidth]{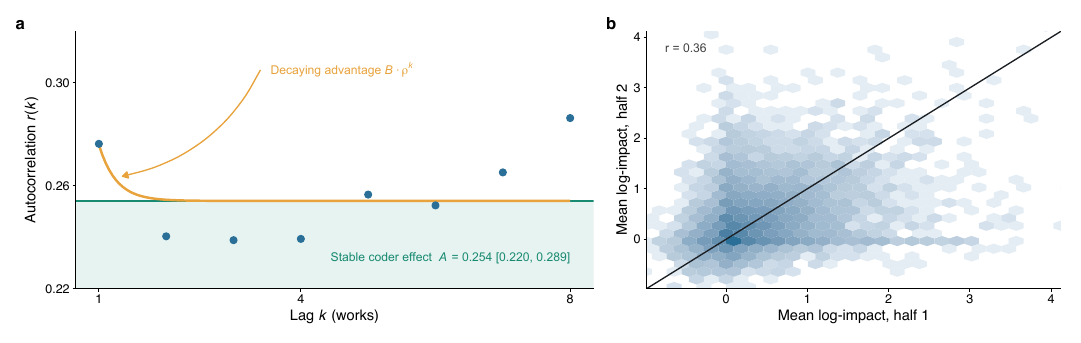}
\caption{\textbf{Decomposition and split-half consistency at the primary condition.}
\textbf{a}, Within-coder autocorrelation of log-impact $k$ works apart (points;
noisy empirical estimates), with the fitted decomposition
$r(k)=\stableshare+B\rho^k$ (amber). The decaying component fades within a
single lag at this condition (small fitted $\rho$), so at every observed lag the fit
is dominated by the flat plateau (teal). That plateau is a correlation floor that
does not fade with distance between works, the signature of a stable coder
effect ($\stableshare$=0.254 [0.220, 0.289]). \textbf{b}, Split-half career
consistency, plotting each coder's mean log-impact in one random half of the career
against the other (hex-binned; $n$=14,091; $r$=0.36).}
\label{fig:skilldecomp}
\end{figure*}

\begin{figure}[t]
\centering
\includegraphics[width=\columnwidth]{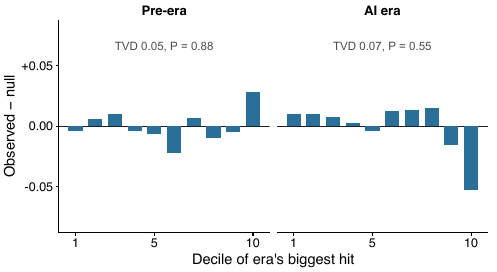}
\caption{\textbf{The lottery is intact in both eras.} Departure of the decile
distribution of each era's biggest hit from that era's own length-mixture
lottery null (held-out real condition, $W$=24), with TVD 0.05 ($p$=0.88) pre-era and 0.07
($p$=0.55) in the AI era; neither rejects. At incumbent sample sizes this is a
power-limited ``no detectable change,'' not a confirmed exact null (see text).}
\label{fig:ailottery}
\end{figure}

\subsection{Identity-Resolution Validation Detail}
\label{app:identity}

We validate identity resolution against a gold set of 28,896 login--identity pairs built from ORCID records, scored separately per era and per world region (Table~\ref{tab:identity}). The design responds to two features of the data. First, GitHub Archive hashes the personal part of each commit author's email address, so authors cannot be matched as strings; instead we decode the hashes with a dictionary of the hashes each account's commits could carry (GitHub's two no-reply address formats, in original and lowercased form, built from the account's login and numeric id), and a decode is either exact or absent. Second, the mix of commits changes over the study window. Platform no-reply addresses grow from 10.1\% of commits in the pre-era to 37.5\% in 2023--25, and continuous-integration services push a growing share of commits authored by someone else. This second shift is what breaks the backup dominance rule of the main text, which assigns an author to the account that pushes most of their commits. In the post-era, roughly 30\% of that rule's assignments would hand human commits to accounts mediated by \texttt{github-actions[bot]}, dragging its raw precision to 0.65--0.69. The bot filter absorbs exactly these cases, restoring precision to $\geq$0.98 in every era and region, while hash-dictionary decoding remains exact by construction. Commits that neither route resolves are left unassigned. Identity error therefore enters recall rather than precision, and an unassigned commit shortens a career rather than mixing two people's work, which biases the analysis against finding a stable factor rather than toward one.

\begin{table}[h]
\centering
\small
\caption{Identity-resolution validation against the ORCID-linked gold set
(28,896 pairs), by era. Regional evaluation (nine world regions) shows the same
pattern; the minimum regional post-filter precision is 0.98.}
\label{tab:identity}
\setlength{\tabcolsep}{3.5pt}
\begin{tabular}{lccc}
\toprule
 & 2015--19 & 2020--22 & 2023--25 \\
\midrule
Commit resolution rate & 60.6\% & 71.8\% & 74.7\% \\
Raw dominance precision & 1.000 & --- & 0.65--0.69 \\
Post-bot-filter precision & 1.000 & $\geq$0.98 & $\geq$0.98 \\
No-reply share of commits & 10.1\% & --- & 37.5\% \\
No-reply decode precision & exact & exact & exact \\
\bottomrule
\end{tabular}
\end{table}

\subsection{Estimator Validation Suite}
\label{app:validation}

Both estimators were validated on synthetic career panels before any contact with the data, against nine generative regimes with twenty pass/fail checks. The suite is adversarial. Each regime simulates a process that a naive estimator would misread as a stable factor, and the estimators must reject it while affirming a genuine one. The regimes include (i) pure cumulative advantage (no person effect; the decomposition must return $\stableshare\approx0$ with the momentum term absorbing the autocorrelation); (ii) near-unit-root autoregression with $\rho$ swept over 0.90--1.02 (the stationarity check must fire); (iii) a drifting person effect (non-stationarity must be flagged rather than fit as a plateau); (iv) measurement-error attacks on the impact proxy (attenuation, not spurious structure); (v) a genuine stable factor of known share (recovered within bootstrap CI); and (vi) zero-inflated variants of the above matching the observed zero share. For the lottery test, the suite also verified that deterministic first-occurrence tie-breaking on zero-heavy proxies manufactures a spurious early lean, and that randomized tie-breaking with all-tie exclusion removes it. That artifact, not any substantive preference, dictated the tie-handling rule that was then frozen in the registration.

\subsection{Full Specification-Grid Results}
\label{app:grid}

\begin{figure*}[t]
\centering
\includegraphics[width=\textwidth]{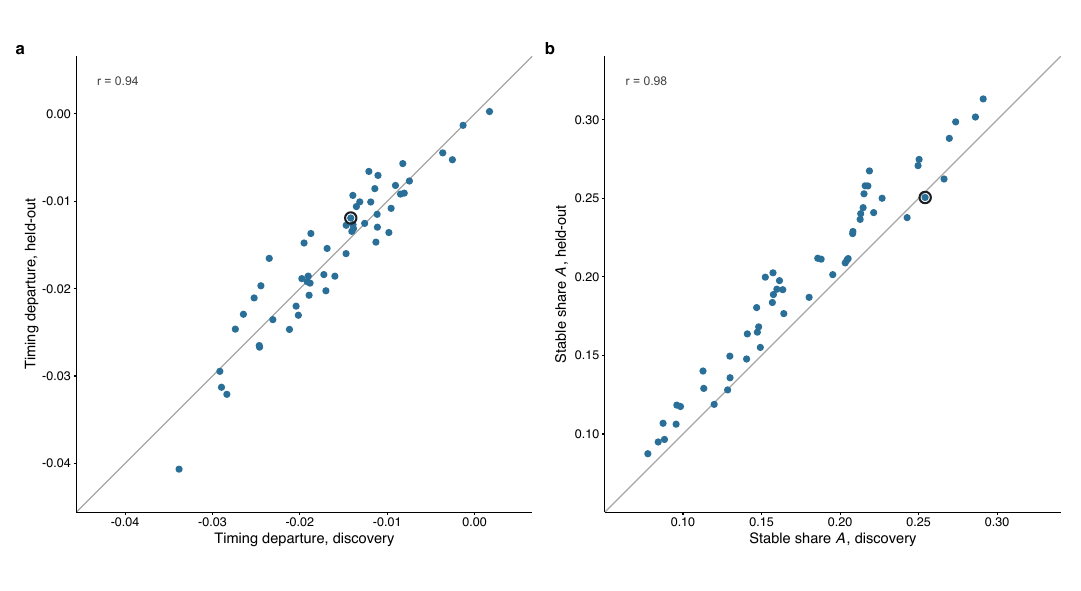}
\caption{\textbf{Held-out confirmation holds condition by condition, not only on average.}
Each point is one of the 54 pre-registered conditions, discovery half
against held-out half; the ringed point is the primary condition. \textbf{a}, Timing
departure (mean biggest-hit position minus lottery expectation), $r$=0.94.
\textbf{b}, Stable share $\stableshare$, $r$=0.98, sitting slightly above the
diagonal (held-out medians marginally larger). The whole grid, not just its
average, carries to the sealed half.}
\label{fig:heldout}
\end{figure*}

Table~\ref{tab:ed1} summarizes RQ1 and RQ2 across both halves. Tables~\ref{tab:a1} and \ref{tab:a2} report every live condition of the pre-registered grid, in the discovery and held-out halves respectively. Each row lists the sample size, the lottery departure (TVD), the stable share with its bootstrap CI, the persistence ratio, the stationarity flag, and the full pre-registered verdict; every condition rejects the exact lottery null at BH-adjusted permutation $p=6.7\times10^{-4}$. The moderator medians discussed in the main text (Table~\ref{tab:mod}) are computed over these rows. $W$=120 conditions are empty at the calendar close (every horizon is right-censored) and are not part of the live grid. Fig.~\ref{fig:heldout} shows, condition by condition, how the discovery results carry to the held-out half, and Fig.~\ref{fig:pipeline} the v1$\,\to\,$v2 pipeline comparison (Appendix~\ref{app:pipeline}).

\begin{table}[t]
\centering
\scriptsize
\caption{Results for RQ1 and RQ2, discovery vs.\ held-out halves, at the
primary condition and across all 54 conditions, plus the calendar sensitivity
checks. C1 counts conditions where the lottery test rejects after correction.
``Verdict'' is the pre-registered three-criterion rule for a meaningful stable
factor: $\stableshare>0.10$ with bootstrap lower bound above 0.02, persistence
ratio $>0.20$, and growth ratio $<$1.5. The dagger marks analyses added after
the pre-registered plan (Appendix~\ref{app:prereg}).}
\label{tab:ed1}
\setlength{\tabcolsep}{3pt}
\resizebox{\columnwidth}{!}{%
\begin{tabular}{lcc}
\toprule
 & Discovery & Held-out \\
\midrule
Careers $n$ (in test) & 14{,}091 (12{,}270) & 13{,}957 (12{,}227) \\
Mean peak pos.\ (null) & 0.548 (0.562) & 0.550 (0.562) \\
TVD; perm.\ $p$ & 0.021; $2{\times}10^{-4}$ & 0.022; $2{\times}10^{-4}$ \\
Grid: BH-reject; med.\ TVD & 54/54; 0.028 & 54/54; 0.029 \\
\midrule
$A$ [95\% CI] & 0.254 [0.220, 0.289] & 0.251 [0.209, 0.292] \\
$B$; ratio $A/(A{+}B)$ & 1.09; 0.189 & 1.14; 0.180 \\
Growth ratio (limit 1.5) & 1.42 & 1.75 \\
Split-half $r$ & 0.363 & 0.351 \\
Grid: CI$>$0; verdict & 54/54; 36/54 & 54/54; 33/54 \\
Grid: median $A$ & 0.172 & 0.201 \\
\midrule
Clean-cadence: C1; verdict & 18/18; 15/18 & 18/18; 13/18$^{\dagger}$ \\
\quad primary-adj.\ $A$; $B$; ratio & 0.310; 0; 1.00 & 0.266; 0.50; 0.35$^{\dagger}$ \\
No-gap: C1; verdict & 54/54; 42/54 & 54/54; 38/54$^{\dagger}$ \\
\quad primary $A$; $B$; ratio & 0.310; 0; 1.00 & 0.267; 0.50; 0.35$^{\dagger}$ \\
Pipeline v1$\to$v2 & \multicolumn{2}{c}{med.\ $\Delta$TVD $-0.010$; med.\ $\Delta A$ $+0.008$} \\
\bottomrule
\end{tabular}}%
\end{table}

\begin{table}[t]
\centering
\scriptsize
\caption{Grid moderators (discovery half; descriptive reading, medians over
conditions sharing each level). The left block
reports the early lean
(median shift of mean hit position from its lottery expectation; median TVD),
the right block the stable share (median $\stableshare$; median split-half $r$). Held-out
medians preserve every ordering discussed in the text.}
\label{tab:mod}
\begin{tabular}{llcccc}
\toprule
Axis & Level & Shift & TVD & $\stableshare$ & Split-half $r$ \\
\midrule
\multirow{3}{*}{Floor} & $\geq$3 & $-0.011$ & 0.019 & 0.182 & 0.288 \\
 & $\geq$5 & $-0.014$ & 0.023 & 0.178 & 0.310 \\
 & $\geq$10 & $-0.024$ & 0.041 & 0.172 & 0.528 \\
\midrule
\multirow{3}{*}{Credit} & R-share & $-0.019$ & 0.031 & 0.235 & 0.361 \\
 & R-lead & $-0.018$ & 0.029 & 0.182 & 0.310 \\
 & R-core & $-0.013$ & 0.023 & 0.117 & 0.262 \\
\midrule
\multirow{2}{*}{Proxy} & reach$^2$ & $-0.014$ & 0.025 & 0.215 & 0.353 \\
 & in-degree & $-0.017$ & 0.029 & 0.157 & 0.316 \\
\midrule
\multirow{3}{*}{$W$ (mo.)} & 24 & $-0.014$ & 0.023 & 0.176 & 0.346 \\
 & 36 & $-0.019$ & 0.033 & 0.175 & 0.349 \\
 & 60 & $-0.012$ & 0.022 & 0.165 & 0.313 \\
\bottomrule
\end{tabular}
\end{table}

\subsection{Pipeline-Replacement Robustness}
\label{app:pipeline}

\begin{figure*}[t]
\centering
\includegraphics[width=\textwidth]{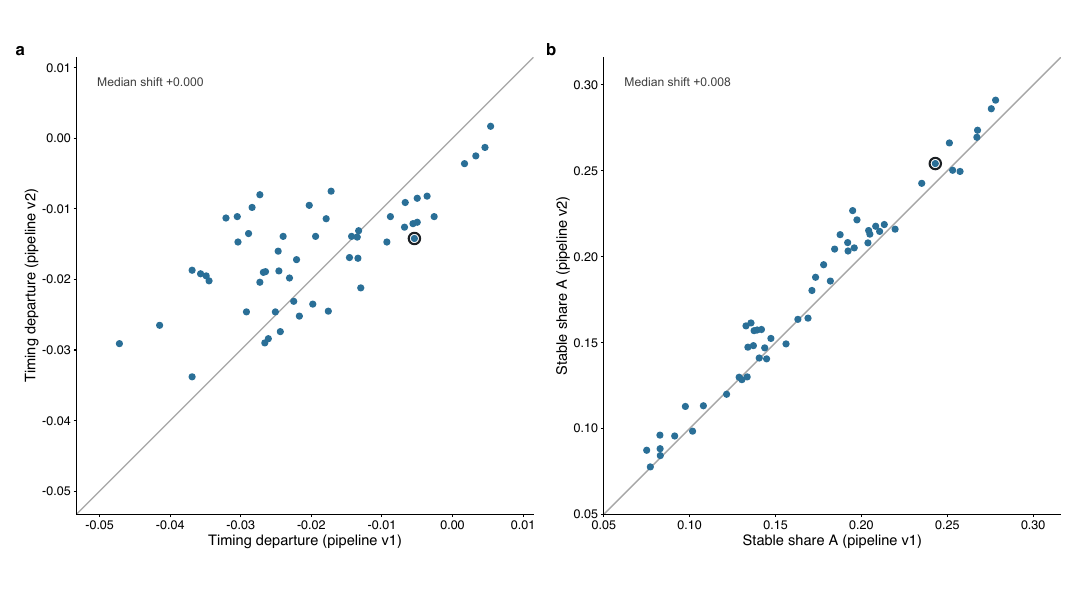}
\caption{\textbf{The pipeline replacement moves no conclusion.} Every grid condition
estimated under the registered v1 pipeline against the final v2 pipeline
(discovery half). \textbf{a}, Timing departure (median shift $+$0.000).
\textbf{b}, Stable share $\stableshare$ (median shift $+$0.008). The ringed
primary condition's stable share moves by less than a sixth of its CI width; its
timing departure moves within the cross-grid spread, and its rejection and
verdict are unchanged.}
\label{fig:pipeline}
\end{figure*}

Between registration and the final analysis, the measurement pipeline was upgraded end to end. Identity resolution moved to v2 (hash-dictionary decoding of no-reply commits, era-local validation), and a panel rebuild enlarged the underlying work table by a factor of $\sim$5.5. If the findings were artifacts of pipeline choices, this replacement should move them. We therefore re-ran the full discovery analysis on both pipeline versions and matched all 54 conditions. The story does not move. Median TVD falls from 0.043 (v1) to 0.028 (v2), so the lottery holds slightly \emph{better} under the improved identity resolution (median per-condition $\Delta$TVD $-0.010$). The stable share is essentially unchanged (median $\stableshare$ 0.170 v1 vs.\ 0.172 v2, median per-condition $\Delta\stableshare$ $+0.008$), and the verdict count moves from 37/54 to 36/54. A result that survives a full replacement of its measurement pipeline with deltas this small is unlikely to be an artifact of either pipeline (Fig.~\ref{fig:pipeline}).

\subsection{Full Pre/Post Grid}
\label{app:prepost}

\begin{figure*}[t]
\centering
\includegraphics[width=\textwidth]{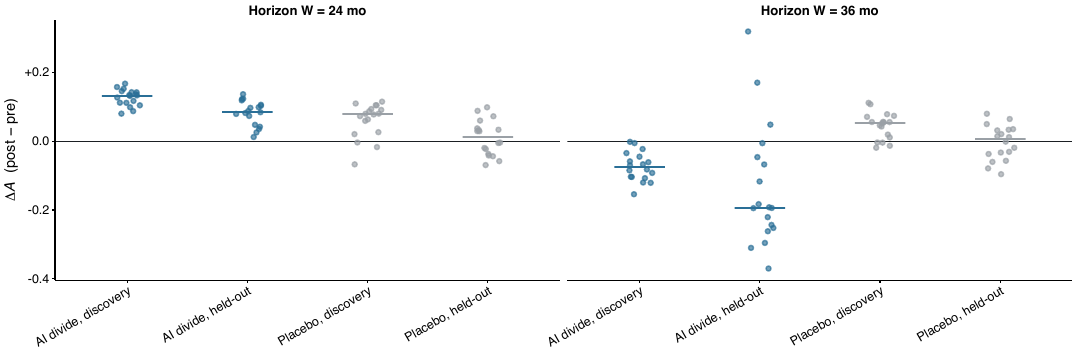}
\caption{\textbf{The full pre/post grid, with the real-divide change inside
the placebo range at the horizon that carries the comparison.} Pre-to-post change
in the stable share for every live
pre/post condition (points), grouped by divide and half, with medians (bars). At
$W$=24 the real-divide change sits within the placebo (aging-only) range in both
halves. At $W$=36, where post-era right-censoring reaches 92--93\%, the
real-divide conditions flip sign between halves while the uncensored placebo
stays near zero; that instability is a symptom of the censoring, not an era effect.}
\label{fig:prepostgrid}
\end{figure*}

Tables~\ref{tab:a3}--\ref{tab:a4b} report all 36 real-divide and 36 placebo-divide conditions per half (one table per divide and half), listing incumbent counts, per-era stable shares, the pre-to-post change, the cross-divide rank correlation of per-coder effects, and the post-era censoring share. The censoring column shows, condition by condition, the imbalance behind the main-text caution. Real-divide $W$=36 conditions carry 92--93\% censoring and small completed-work counts, and their $\Delta\stableshare$ is sign-unstable across halves, while $W$=24 conditions (censoring $\approx$0.5) and all placebo conditions (censoring 0) behave stably. Fig.~\ref{fig:prepostgrid} shows the full $\Delta\stableshare$ distributions by horizon.

\begin{table}[t]
\centering
\scriptsize
\caption{RQ3 at the primary pre/post condition: the AI divide vs.\ the aging
placebo. $\rho$ is the Spearman rank correlation of per-coder effects across
the divide. Held-out real-divide per-era 95\% CIs are $A_{\text{pre}}$
[0.206, 0.368] and $A_{\text{post}}$ [0.243, 0.464]; placebo [0.094, 0.390]
and [0.203, 0.468]. The dagger marks analyses added after the pre-registered
plan, computed on the paired subset (Appendix~\ref{app:prereg}); the paired
real and placebo $\Delta A$ (+0.087 and +0.052 held-out) differ slightly from
the all-incumbent rows above. Across all 36 conditions, the real median
$\Delta A$ is +0.039 in both halves; at $W$=24, real +0.086 (held-out) vs.\
placebo +0.012, while $W$=36 is unstable under censoring (see text). No
real-divide timing condition rejects after correction in either half.}
\label{tab:ed2}
\setlength{\tabcolsep}{3pt}
\begin{tabular}{llccccc}
\toprule
Half & Divide & $n$ & $A_{\text{pre}}$ & $A_{\text{post}}$ & $\Delta A$ & $\rho$ \\
\midrule
Disc. & Real & 412 & 0.291 & 0.403 & $+0.112$ & 0.361 \\
Disc. & Placebo & 199 & 0.238 & 0.342 & $+0.105$ & 0.490 \\
\textbf{Held-out} & \textbf{Real} & \textbf{421} & \textbf{0.294} & \textbf{0.381} & $\bm{+0.088}$ & \textbf{0.303} \\
Held-out & Placebo & 217 & 0.306 & 0.366 & $+0.060$ & 0.399 \\
\midrule
\multicolumn{7}{l}{DiD $D^{\dagger}$: $+0.035$ $[-0.218, +0.230]$ h-o; $+0.016$ $[-0.293, +0.396]$ disc.} \\
\multicolumn{7}{l}{\quad TOST $\pm$0.10 fails (0/36 conditions); all 72 $D$ CIs cover 0} \\
\multicolumn{7}{l}{Matched-cens.\ placebo$^{\dagger}$: $\Delta A$ $+0.176$ h-o, $+0.218$ disc.} \\
\multicolumn{7}{l}{\quad $D$ $-0.089$ $[-0.540, +0.340]$ h-o; $-0.107$ $[-0.482, +0.258]$ disc.} \\
\multicolumn{7}{l}{Career-clock$^{\dagger}$: $\Delta A$ $+0.000$ $[-0.247, +0.271]$ h-o;} \\
\multicolumn{7}{l}{\quad $+0.108$ $[-0.237, +0.553]$ disc.} \\
\bottomrule
\end{tabular}
\end{table}

\paragraph{Direct era-contrast test and censoring parity (post-unsealing).}
Both analyses were specified in a dated note before any of their numbers were computed (Appendix~\ref{app:prereg}) and run with the frozen estimators. The difference-in-differences contrast $D$ is estimated by a joint bootstrap that resamples contributors with their pre and post sequences kept paired, independently within the real and placebo cohorts ($B$=1{,}000 at the primary condition, 200 elsewhere; per-condition seeds recorded in the outputs). It is computed on the paired subset of contributors with realized works in both eras, which is why its per-era values can differ slightly from the all-incumbent values of Tables~\ref{tab:a3}--\ref{tab:a4b}. At the primary condition, $D$=+0.035 $[-0.218, +0.230]$ held-out and +0.016 $[-0.293, +0.396]$ discovery. All 72 condition-level $D$ CIs across both halves cover zero (medians +0.026 and $-0.008$), and the $\pm$0.10 TOST equivalence test passes in 0 of 72, so the data rule out large era shifts but cannot certify that small ones are absent. The timing interaction $T$, the pre-to-post change in TVD for the real divide minus the placebo, is +0.005 $[-0.080, +0.117]$ held-out. All 36 held-out CIs cover zero; four discovery in-degree conditions (each at the $\geq$10 floor or in the $W$=36 column) show positive $T$ CIs excluding zero, and none replicates held-out.

The matched-censoring placebo imposes the real administrative censoring rule on the placebo cohort at a pseudo-freeze of April 2020. The calendar mirror censors 61--64\% of placebo pseudo-post rows at $W$=24, stricter than the real snapshot-based 49--52\%, and censors all of them at $W$=36, where the window is shorter than the horizon; the matched grid is therefore the 18 $W$=24 conditions. Equalized censoring raises the placebo $\Delta\stableshare$ to a grid median of +0.084 held-out and +0.110 discovery (primary condition +0.176 and +0.218), above the corresponding real values, and the primary-condition contrast becomes $D$=$-0.089$ $[-0.540, +0.340]$ held-out ($-0.107$ $[-0.482, +0.258]$ discovery). In other words, restricting to fully measured works inflates apparent pre-to-post gains, and once the placebo carries the same inflation no residual remains for an era effect. The specification note anticipated two outcomes, the matched placebo moving toward the real divide (indicating a censoring artifact) or away from it (indicating that censoring had masked a real effect). The realized outcome falls outside that dichotomy. The matched placebo moved toward and \emph{past} the real divide, so $|D|$ grew while changing sign. We interpret this as supporting the no-break conclusion, because the masking scenario requires the equalized placebo to sit below the real divide, and a negative point estimate weighs all the more against a positive era effect. The matched CI is wide, however, and this interpretation rests on the point estimates.

\paragraph{Remaining registered aging controls and convergent validity (post-unsealing).}
A second dated note (2026-07-19) specified the amendment's two remaining aging controls and the registered convergent-validity statistic in full before any of their numbers were computed; all three ran on both halves with the frozen modules imported read-only.

The career-clock alignment replaces each incumbent's whole pre-era career with their last $\min(n_{\text{pre}}, n_{\text{post}})$ completed pre-era works, ordered as in the frozen run, so that the pre and post sequences sit at matched career position; qualification, censoring, normalization and estimators are otherwise identical. At the primary condition, the held-out matched sequences give a stable share of 0.381 late-pre and 0.381 post ($\Delta\stableshare$=+0.000 $[-0.247, +0.271]$; discovery +0.108 $[-0.237, +0.553]$), and the per-person ordering carries between the matched sequences at rank $\rho$=0.27 held-out (0.42 discovery). The discovery point estimate sits marginally above the specification note's +0.10 materiality threshold for a masked era shift, but its CI is wide and covers zero, and the note's fixed decision rule keys off the held-out half, which comes in at +0.000. The no-break conclusion therefore leans on the confirmatory half here, as it does for the RQ2 composite verdict. Across the grid, condition CIs cover zero in 36/36 discovery and 32/36 held-out, with sign-unstable cross-half medians (+0.138 discovery, $-0.172$ held-out).

The career-age covariate control stratifies each era-local length-mixture null into terciles of career age at era entry. Because the exact null already conditions on each contributor's own career length, the aggregated stratified expectation coincides with the unstratified one. Empirically, every one of the 288 era-condition TVDs is unchanged ($|\Delta|<10^{-4}$), and the stratified permutation test moves a raw $\alpha$=0.05 conclusion in four boundary cases ($p$ crossing 0.05 by $\leq$0.007), none affecting an FDR conclusion. Hit position correlates only weakly with career age at era entry (primary condition $\rho$ between $-0.16$ and $+0.10$ across eras and halves; the significant correlations concentrate in pre-era ages and are negative, consistent with the early lean growing with career length). The placebo runs collapse to a single stratum by construction, because the 2015--16 entry band leaves too little age variation to tercile, and are flagged as such in the outputs.

The convergent-validity statistic recomputes per-person effects at the main-grid primary condition separately under reach$^2$ and under in-degree, on the identical contributor roster (qualification counts works, not impact values, so the intersection is each half's full roster), and correlates them. Spearman $\rho$=0.979 in both halves ($n$=14{,}091 discovery, 13{,}957 held-out) clears the registered $\geq$0.7 threshold, so under the registered rule the two proxies may be treated as measuring one factor.

\subsection{Calendar-Gap Sensitivity in Full}
\label{app:sensitivity}

\begin{figure*}[t]
\centering
\includegraphics[width=\textwidth]{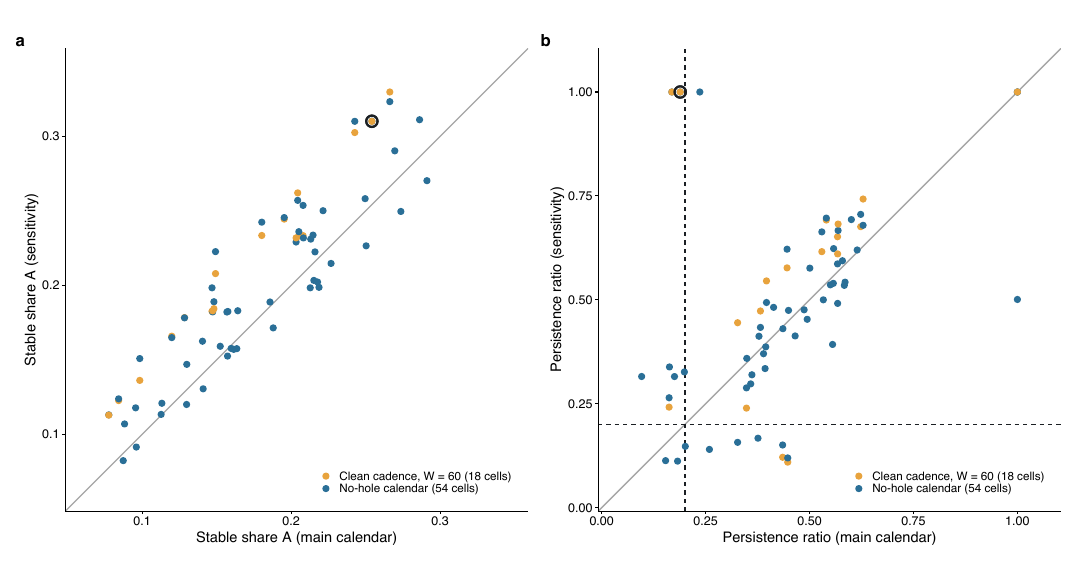}
\caption{\textbf{Calendar sensitivities systematically raise the stable share
and, at the verdict-critical conditions, the persistence ratio (discovery half).} Each
condition of the pre-registered grid under
the main calendar
(x) against the two calendar-sensitivity variants (y), no-gap (blue, 54 conditions) and
clean-cadence (amber, 18 conditions at $W$=60). \textbf{a}, Stable share
$\stableshare$, where nearly all conditions sit above the diagonal. \textbf{b}, Persistence
ratio against the 0.20 threshold (dashed); the $W$=60 conditions near the threshold under the
main
calendar, including the ringed primary(-adjacent) condition, move to ratio 1.00 when
the snapshot-gap artifact is removed. The held-out post-unsealing re-run
replicates the direction but not the full reversal (ratio 0.18 to 0.35,
$B$ halved but positive; Table~\ref{tab:ed1} and
Appendix~\ref{app:sensitivity}).}
\label{fig:calsens}
\end{figure*}

Table~\ref{tab:a5} reports the clean-cadence subgroup in full (18 conditions; careers beginning August 2017 or later, whose $W$=60 horizons are fully covered by dense snapshots). The no-gap variant instead keeps all careers and drops the horizon rows that cross the January 2020 to August 2022 snapshot gap. At the primary condition the no-gap variant gives $\stableshare$=0.310 [0.244, 0.364], $B$=0, ratio 1.00, in agreement with the clean-cadence estimate at the primary-adjacent condition ($\stableshare$=0.310 [0.240, 0.376], $B$=0, ratio 1.00). The two analyses remove the calendar artifact by different mechanisms, sample restriction versus row exclusion, and produce the same reversal of the primary condition's verdict failure. Fig.~\ref{fig:calsens} compares both variants to the main calendar condition by condition (discovery half).

Both sensitivities were subsequently re-run on the held-out half. This happened after the unseal and after the held-out primary estimates were read, so we label these runs post-unsealing analyses; the code was byte-identical to the discovery-half sensitivity runs. The held-out half partially replicates the discovery pattern. The two variants again agree with each other almost exactly at the primary condition, with no-gap giving $\stableshare$=0.267 [0.218, 0.318], $B$=0.504, ratio 0.346, and clean-cadence $\stableshare$=0.266 [0.223, 0.307], $B$=0.502, ratio 0.346. Relative to the held-out main calendar (ratio 0.180, $B$=1.14), the ratio again rises and clears the 0.20 threshold and $B$ again falls, but the discovery half's full collapse to $B$=0, ratio 1.00 does not reproduce, and the composite verdict still fails at the stationarity limit (growth ratio 1.53 vs.\ 1.5). Grid-wide, the held-out sensitivity runs give C1 rejections in every condition (54/54 no-gap, 18/18 clean-cadence) with verdict passes in 38/54 and 13/18 conditions respectively (median $\stableshare$ 0.221 and 0.181). These held-out numbers are why we treat the direction of the calendar effect as replicated. Its magnitude, and with it the exact $A$-vs-$B$ division, we treat as less securely identified (main text, RQ2).

\subsection{Preregistration Timeline and Deviations}
\label{app:prereg}

The analysis plan (grid, estimators, split salt, C1 tie handling, C2 verdict thresholds) was registered before data collection. The pre/post (RQ3) specification, including the incumbent definition, the common-horizon rule, the placebo design and a three-outcome interpretation table, was frozen by a timestamped amendment after the pre-era discovery read but before any post-era estimate existed. The same amendment consolidated the registered multiverse to the 54 reported conditions and fixed the era-comparison horizons. The pre/post readout horizon ($W$=24 as primary, $W$=36 as diagnostic) was fixed in the analysis code at the discovery stage, before the unseal, and is not named in the amendment itself. Because the amendment was posted after the discovery grid had been read, the consolidation is itself a post-discovery decision, applied identically to both halves.

Relative to the original registration, the consolidation dropped three secondary impact proxies (forks, downloads, engagement), the stars negative control, the commit-level granularity check, and ecosystem stratification; the registered dependent-count proxy is retained as in-degree. None of the dropped specifications appears in any sealed output, so no estimated condition was omitted. Two registered normalization choices changed at implementation. The registered primary normalization stratum (ecosystem, cohort-year, language) was not implementable because language recovery was unavailable at scale, so the registered alternative (ecosystem, cohort-year) is used for all conditions, a deviation on the primary specification. The registered primary proxy, transitive downstream reach, is operationalized as its two-hop truncation (reach$^2$) on tractability grounds; the change is documented in the build specification and carried into the consolidating amendment, which names reach$^2$ in the primary condition. Two registered channels, antecedents (C3) and measurement bias (C4, including its region-stratified gold-set validation, whose roster floors were met), are deferred in full rather than reported here; region-level statements in this paper are accordingly exploratory only (Limitations).

Two further implementation deviations are on record. The $\rho$ grid of the Eq.~1 fit runs to 0.9 rather than the registered 0.98. On the 54-condition grid this cap never binds (max fitted $\rho$=0.68 discovery, 0.58 held-out; the calendar sensitivities reach 0.76, also non-binding). In the era-local pre/post fits it does bind, with $\rho$ pinned at 0.9 in 25/72 (real, discovery), 11/72 (real, held-out), and 11/72 and 8/72 (placebo) era-fits; the real-divide pins concentrate in the heavily censored post-era $W$=36 column, and the placebo pins scatter across small-$n$ pre-era conditions. No headline number derives from a cap-pinned fit; the real-divide primary condition fits $\rho\leq0.40$ pre and 0.0 post in both halves, and the placebo primary reaches 0.74, also unpinned. Separately, of the amendment's three specified aging controls, only the placebo divide ran before the unseal. The other two (the career-clock alignment and the career-age covariate variant), together with the registered cross-proxy convergent-validity statistic, which could not be computed from the sealed outputs because per-person effects per proxy were not retained there, were specified in full in the second dated note (2026-07-19) and then run on both halves with the frozen modules imported read-only. They are reported as labeled post-unsealing analyses with the registered decision rules applied as registered (RQ2; RQ3; Appendix~\ref{app:prepost}). All three support the sealed conclusions, and the convergent-validity correlation (0.979 in both halves) clears the registered 0.7 threshold for treating the two proxies as one factor.

The held-out half was unsealed once, on record (2026-07-12, UTC-timestamped attestation), with analysis code byte-identical to the discovery run. No held-out number was computed before the unseal, and no sealed number was recomputed after. Three sets of additional computations post-date the unseal. First, the two calendar sensitivities were re-run on the held-out half after the held-out primary estimates were read, with code byte-identical to the discovery-half sensitivity runs. Second, in a pre-submission robustness pass, a direct difference-in-differences test of the era contrast (paired-bootstrap $D$ and timing interaction $T$, with a $\pm$0.10 TOST margin fixed at the registration's $\stableshare$ materiality threshold) and a matched-censoring placebo were specified in full in a dated note (2026-07-12, before any of their numbers were computed) and then run on both halves with the frozen estimators. Third, the remaining registered analyses (the career-clock alignment, the career-age covariate variant of the timing null, and the cross-proxy convergent-validity statistic) were specified in a second dated note (2026-07-19), again before any of their numbers were computed, and run on both halves with the frozen modules imported read-only. All of these are labeled post-unsealing analyses wherever they appear (RQ2; RQ3; Tables~\ref{tab:ed1}--\ref{tab:ed2}; Appendices~\ref{app:prepost} and \ref{app:sensitivity}) and are not treated as sealed confirmation. One code deviation is on record. An earlier qualification routine incorrectly applied estimator-stage censoring to incumbent qualification, structurally excluding every post-era incumbent (0 in every condition). It was detected and fixed at the discovery stage, before any real post-era estimate existed; the buggy outputs are archived and the fix is documented in a public deviation note. Grid-moderator and pipeline-robustness summaries (Table~\ref{tab:mod}; the v1$\to$v2 comparison) are descriptive readings of pre-registered quantities, not pre-registered hypotheses, and are labeled as such wherever they appear.

% Auto-generated from sealed results_v2 JSONs — do not hand-edit numbers.

\begin{table*}[p]
\centering
\scriptsize
\setlength{\tabcolsep}{4.5pt}
\caption{Full pre-registered specification grid, discovery half (all 54 live conditions; * marks the
pre-registered primary condition). All conditions reject the exact lottery null under BH-FDR
(adjusted permutation $p=6.7\times10^{-4}$ in every condition). Stat.\ marks variance stationarity (growth $<$1.5) and Verdict the full pre-registered stable-factor rule.}
\label{tab:a1}
\begin{tabular}{llllrcccccc}
\toprule
Credit & $W$ & Floor & Proxy & $n$ & TVD & $A$ & 95\% CI & Ratio & Stat. & Verdict \\
\midrule
R-core & 24 & 3 & deg & 33,975 & 0.018 & 0.087 & [0.07, 0.10] & 0.36 & T & F \\
R-core & 24 & 3 & r$^2$ & 33,975 & 0.019 & 0.130 & [0.11, 0.15] & 0.35 & T & T \\
R-core & 24 & 5 & deg & 19,176 & 0.023 & 0.096 & [0.08, 0.11] & 0.39 & T & F \\
R-core & 24 & 5 & r$^2$ & 19,176 & 0.023 & 0.141 & [0.12, 0.17] & 0.39 & T & T \\
R-core & 24 & 10 & deg & 6,324 & 0.036 & 0.113 & [0.10, 0.13] & 0.44 & T & T \\
R-core & 24 & 10 & r$^2$ & 6,324 & 0.038 & 0.164 & [0.14, 0.19] & 0.49 & T & T \\
R-core & 36 & 3 & deg & 30,768 & 0.023 & 0.088 & [0.08, 0.10] & 0.38 & T & F \\
R-core & 36 & 3 & r$^2$ & 30,768 & 0.023 & 0.130 & [0.11, 0.15] & 0.38 & T & T \\
R-core & 36 & 5 & deg & 17,792 & 0.029 & 0.096 & [0.08, 0.11] & 0.41 & T & F \\
R-core & 36 & 5 & r$^2$ & 17,792 & 0.025 & 0.140 & [0.12, 0.16] & 0.39 & T & T \\
R-core & 36 & 10 & deg & 5,973 & 0.038 & 0.113 & [0.10, 0.13] & 0.45 & T & T \\
R-core & 36 & 10 & r$^2$ & 5,973 & 0.037 & 0.164 & [0.14, 0.20] & 0.50 & T & T \\
R-core & 60 & 3 & deg & 23,780 & 0.019 & 0.078 & [0.07, 0.09] & 0.38 & T & F \\
R-core & 60 & 3 & r$^2$ & 23,780 & 0.017 & 0.120 & [0.10, 0.14] & 0.44 & T & T \\
R-core & 60 & 5 & deg & 14,382 & 0.020 & 0.084 & [0.07, 0.10] & 0.40 & T & F \\
R-core & 60 & 5 & r$^2$ & 14,382 & 0.021 & 0.128 & [0.11, 0.15] & 0.45 & T & T \\
R-core & 60 & 10 & deg & 5,037 & 0.027 & 0.098 & [0.08, 0.11] & 0.45 & T & F \\
R-core & 60 & 10 & r$^2$ & 5,037 & 0.030 & 0.149 & [0.12, 0.19] & 0.54 & T & T \\
R-lead & 24 & 3 & deg & 28,508 & 0.020 & 0.160 & [0.14, 0.17] & 0.47 & T & T \\
R-lead & 24 & 3 & r$^2$ & 28,508 & 0.017 & 0.215 & [0.19, 0.26] & 0.15 & T & F \\
R-lead & 24 & 5 & deg & 16,300 & 0.030 & 0.161 & [0.14, 0.18] & 0.49 & T & T \\
R-lead & 24 & 5 & r$^2$ & 16,300 & 0.022 & 0.218 & [0.19, 0.25] & 0.18 & T & F \\
R-lead & 24 & 10 & deg & 5,501 & 0.045 & 0.157 & [0.13, 0.18] & 0.53 & T & T \\
R-lead & 24 & 10 & r$^2$ & 5,501 & 0.038 & 0.219 & [0.18, 0.25] & 0.58 & T & T \\
R-lead & 36 & 3 & deg & 25,749 & 0.026 & 0.157 & [0.14, 0.17] & 0.56 & T & T \\
R-lead & 36 & 3 & r$^2$ & 25,749 & 0.028 & 0.213 & [0.18, 0.24] & 0.16 & T & F \\
R-lead & 36 & 5 & deg & 15,056 & 0.033 & 0.158 & [0.14, 0.17] & 0.56 & T & T \\
R-lead & 36 & 5 & r$^2$ & 15,056 & 0.033 & 0.215 & [0.19, 0.25] & 0.36 & T & T \\
R-lead & 36 & 10 & deg & 5,144 & 0.046 & 0.152 & [0.12, 0.17] & 0.55 & T & T \\
R-lead & 36 & 10 & r$^2$ & 5,144 & 0.042 & 0.216 & [0.17, 0.25] & 0.61 & T & T \\
R-lead & 60 & 3 & deg & 19,759 & 0.018 & 0.147 & [0.12, 0.17] & 0.57 & T & T \\
R-lead & 60 & 3 & r$^2$ & 19,759 & 0.021 & 0.203 & [0.17, 0.24] & 0.16 & T & F \\
R-lead & 60 & 5 & deg & 12,069 & 0.022 & 0.148 & [0.12, 0.17] & 0.57 & T & T \\
R-lead & 60 & 5 & r$^2$ & 12,069 & 0.022 & 0.205 & [0.17, 0.24] & 0.33 & T & T \\
R-lead & 60 & 10 & deg & 4,255 & 0.045 & 0.147 & [0.12, 0.17] & 0.53 & T & T \\
R-lead & 60 & 10 & r$^2$ & 4,255 & 0.044 & 0.208 & [0.16, 0.24] & 0.62 & T & T \\
R-share & 24 & 3 & deg & 33,280 & 0.019 & 0.227 & [0.21, 0.24] & 0.20 & T & T \\
R-share & 24 & 3 & r$^2$ & 33,280 & 0.019 & 0.291 & [0.26, 0.33] & 1.00 & F & F \\
R-share & 24 & 5 & deg & 18,925 & 0.029 & 0.213 & [0.20, 0.23] & 0.10 & T & F \\
R-share & 24 & 5 & r$^2$ & 18,925 & 0.023 & 0.274 & [0.24, 0.31] & 0.18 & F & F \\
R-share & 24 & 10 & deg & 6,302 & 0.047 & 0.188 & [0.17, 0.21] & 0.56 & T & T \\
R-share & 24 & 10 & r$^2$ & 6,302 & 0.033 & 0.250 & [0.22, 0.28] & 0.59 & F & F \\
R-share & 36 & 3 & deg & 30,029 & 0.032 & 0.221 & [0.20, 0.24] & 0.24 & T & T \\
R-share & 36 & 3 & r$^2$ & 30,029 & 0.030 & 0.286 & [0.26, 0.32] & 1.00 & T & T \\
R-share & 36 & 5 & deg & 17,507 & 0.037 & 0.208 & [0.19, 0.22] & 0.26 & T & T \\
R-share & 36 & 5 & r$^2$ & 17,507 & 0.036 & 0.270 & [0.24, 0.30] & 0.20 & T & F \\
R-share & 36 & 10 & deg & 5,951 & 0.052 & 0.186 & [0.17, 0.21] & 0.58 & T & T \\
R-share & 36 & 10 & r$^2$ & 5,951 & 0.045 & 0.250 & [0.22, 0.28] & 0.60 & T & T \\
R-share & 60 & 3 & deg & 23,114 & 0.017 & 0.204 & [0.18, 0.23] & 0.17 & T & F \\
R-share & 60 & 3 & r$^2$ & 23,114 & 0.018 & 0.266 & [0.24, 0.30] & 1.00 & T & T \\
R-share & 60 & 5 & deg & 14,091 & 0.023 & 0.195 & [0.17, 0.22] & 0.35 & T & T \\
R-share\textbf{*} & 60 & 5 & r$^2$ & 14,091 & 0.021 & 0.254 & [0.22, 0.29] & 0.19 & T & F \\
R-share & 60 & 10 & deg & 4,975 & 0.045 & 0.180 & [0.16, 0.20] & 0.57 & T & T \\
R-share & 60 & 10 & r$^2$ & 4,975 & 0.040 & 0.243 & [0.21, 0.29] & 0.63 & T & T \\
\bottomrule
\end{tabular}
\end{table*}

\begin{table*}[p]
\centering
\scriptsize
\setlength{\tabcolsep}{4.5pt}
\caption{Full pre-registered specification grid, held-out half (all 54 live conditions; * marks the
pre-registered primary condition). All conditions reject the exact lottery null under BH-FDR
(adjusted permutation $p=6.7\times10^{-4}$ in every condition). Stat.\ marks variance stationarity (growth $<$1.5) and Verdict the full pre-registered stable-factor rule.}
\label{tab:a2}
\begin{tabular}{llllrcccccc}
\toprule
Credit & $W$ & Floor & Proxy & $n$ & TVD & $A$ & 95\% CI & Ratio & Stat. & Verdict \\
\midrule
R-core & 24 & 3 & deg & 34,423 & 0.020 & 0.107 & [0.09, 0.13] & 0.39 & T & T \\
R-core & 24 & 3 & r$^2$ & 34,423 & 0.019 & 0.150 & [0.12, 0.18] & 0.28 & T & T \\
R-core & 24 & 5 & deg & 19,507 & 0.025 & 0.118 & [0.09, 0.14] & 0.43 & T & T \\
R-core & 24 & 5 & r$^2$ & 19,507 & 0.023 & 0.164 & [0.13, 0.20] & 0.40 & T & T \\
R-core & 24 & 10 & deg & 6,447 & 0.035 & 0.140 & [0.11, 0.17] & 0.54 & T & T \\
R-core & 24 & 10 & r$^2$ & 6,447 & 0.030 & 0.192 & [0.15, 0.23] & 0.59 & T & T \\
R-core & 36 & 3 & deg & 31,154 & 0.022 & 0.097 & [0.08, 0.12] & 0.35 & T & F \\
R-core & 36 & 3 & r$^2$ & 31,154 & 0.020 & 0.136 & [0.11, 0.16] & 0.29 & T & T \\
R-core & 36 & 5 & deg & 18,034 & 0.031 & 0.106 & [0.09, 0.13] & 0.41 & T & T \\
R-core & 36 & 5 & r$^2$ & 18,034 & 0.029 & 0.148 & [0.12, 0.18] & 0.38 & T & T \\
R-core & 36 & 10 & deg & 6,107 & 0.041 & 0.129 & [0.10, 0.16] & 0.54 & T & T \\
R-core & 36 & 10 & r$^2$ & 6,107 & 0.037 & 0.177 & [0.14, 0.22] & 0.58 & T & T \\
R-core & 60 & 3 & deg & 23,641 & 0.019 & 0.087 & [0.07, 0.11] & 0.31 & T & F \\
R-core & 60 & 3 & r$^2$ & 23,641 & 0.019 & 0.119 & [0.09, 0.16] & 0.28 & T & T \\
R-core & 60 & 5 & deg & 14,262 & 0.025 & 0.095 & [0.08, 0.12] & 0.36 & T & F \\
R-core & 60 & 5 & r$^2$ & 14,262 & 0.024 & 0.128 & [0.10, 0.17] & 0.34 & T & T \\
R-core & 60 & 10 & deg & 5,017 & 0.035 & 0.118 & [0.09, 0.15] & 0.49 & T & T \\
R-core & 60 & 10 & r$^2$ & 5,017 & 0.035 & 0.155 & [0.12, 0.21] & 0.51 & T & T \\
R-lead & 24 & 3 & deg & 28,847 & 0.021 & 0.192 & [0.16, 0.23] & 0.19 & T & F \\
R-lead & 24 & 3 & r$^2$ & 28,847 & 0.019 & 0.253 & [0.21, 0.30] & 1.00 & T & T \\
R-lead & 24 & 5 & deg & 16,571 & 0.021 & 0.198 & [0.16, 0.24] & 0.22 & T & T \\
R-lead & 24 & 5 & r$^2$ & 16,571 & 0.023 & 0.258 & [0.22, 0.31] & 0.23 & T & T \\
R-lead & 24 & 10 & deg & 5,579 & 0.036 & 0.203 & [0.16, 0.25] & 0.65 & T & T \\
R-lead & 24 & 10 & r$^2$ & 5,579 & 0.029 & 0.267 & [0.21, 0.33] & 0.67 & T & T \\
R-lead & 36 & 3 & deg & 26,031 & 0.033 & 0.184 & [0.16, 0.22] & 0.17 & T & F \\
R-lead & 36 & 3 & r$^2$ & 26,031 & 0.032 & 0.240 & [0.21, 0.29] & 0.54 & T & T \\
R-lead & 36 & 5 & deg & 15,234 & 0.036 & 0.189 & [0.16, 0.23] & 0.13 & T & F \\
R-lead & 36 & 5 & r$^2$ & 15,234 & 0.033 & 0.244 & [0.20, 0.30] & 0.21 & T & T \\
R-lead & 36 & 10 & deg & 5,268 & 0.050 & 0.200 & [0.16, 0.24] & 0.66 & T & T \\
R-lead & 36 & 10 & r$^2$ & 5,268 & 0.043 & 0.258 & [0.21, 0.31] & 0.65 & T & T \\
R-lead & 60 & 3 & deg & 19,549 & 0.019 & 0.165 & [0.13, 0.20] & 0.14 & T & F \\
R-lead & 60 & 3 & r$^2$ & 19,549 & 0.020 & 0.209 & [0.17, 0.26] & 0.25 & T & T \\
R-lead & 60 & 5 & deg & 11,901 & 0.026 & 0.168 & [0.13, 0.21] & 0.42 & T & T \\
R-lead & 60 & 5 & r$^2$ & 11,901 & 0.026 & 0.212 & [0.17, 0.27] & 0.15 & T & F \\
R-lead & 60 & 10 & deg & 4,290 & 0.036 & 0.180 & [0.14, 0.23] & 0.64 & T & T \\
R-lead & 60 & 10 & r$^2$ & 4,290 & 0.033 & 0.228 & [0.18, 0.27] & 0.64 & F & F \\
R-share & 24 & 3 & deg & 33,636 & 0.024 & 0.250 & [0.23, 0.27] & 1.00 & T & T \\
R-share & 24 & 3 & r$^2$ & 33,636 & 0.025 & 0.313 & [0.27, 0.35] & 1.00 & F & F \\
R-share & 24 & 5 & deg & 19,250 & 0.029 & 0.237 & [0.21, 0.27] & 0.19 & T & F \\
R-share & 24 & 5 & r$^2$ & 19,250 & 0.027 & 0.299 & [0.26, 0.34] & 0.47 & F & F \\
R-share & 24 & 10 & deg & 6,443 & 0.044 & 0.211 & [0.18, 0.24] & 0.66 & T & T \\
R-share & 24 & 10 & r$^2$ & 6,443 & 0.037 & 0.275 & [0.24, 0.31] & 0.67 & F & F \\
R-share & 36 & 3 & deg & 30,423 & 0.031 & 0.241 & [0.21, 0.27] & 1.00 & T & T \\
R-share & 36 & 3 & r$^2$ & 30,423 & 0.032 & 0.302 & [0.27, 0.34] & 1.00 & F & F \\
R-share & 36 & 5 & deg & 17,757 & 0.041 & 0.229 & [0.21, 0.26] & 0.18 & T & F \\
R-share & 36 & 5 & r$^2$ & 17,757 & 0.038 & 0.288 & [0.24, 0.33] & 0.37 & F & F \\
R-share & 36 & 10 & deg & 6,078 & 0.059 & 0.212 & [0.18, 0.24] & 0.67 & T & T \\
R-share & 36 & 10 & r$^2$ & 6,078 & 0.049 & 0.271 & [0.23, 0.30] & 0.67 & F & F \\
R-share & 60 & 3 & deg & 22,916 & 0.018 & 0.211 & [0.18, 0.24] & 0.25 & F & F \\
R-share & 60 & 3 & r$^2$ & 22,916 & 0.018 & 0.262 & [0.22, 0.31] & 0.83 & F & F \\
R-share & 60 & 5 & deg & 13,957 & 0.023 & 0.201 & [0.17, 0.24] & 0.41 & T & T \\
R-share\textbf{*} & 60 & 5 & r$^2$ & 13,957 & 0.022 & 0.251 & [0.21, 0.29] & 0.18 & F & F \\
R-share & 60 & 10 & deg & 5,012 & 0.045 & 0.187 & [0.15, 0.22] & 0.62 & T & T \\
R-share & 60 & 10 & r$^2$ & 5,012 & 0.037 & 0.238 & [0.19, 0.28] & 0.63 & F & F \\
\bottomrule
\end{tabular}
\end{table*}

\begin{table*}[p]
\centering
\scriptsize
\setlength{\tabcolsep}{4.5pt}
\caption{Real divide, discovery half (all 36 pre/post conditions; * marks the pre-registered
primary pre/post condition). $\rho$ is the Spearman rank correlation of per-coder effects across the divide, and Cens.\ the share of post-era rows right-censored at the estimator stage.
Real-divide $W$=36 conditions carry heavy censoring and are reported as a diagnostic, not an estimate (main text).}
\label{tab:a3}
\begin{tabular}{llllrccccc}
\toprule
Credit & $W$ & Floor & Proxy & $n$ & $A_{\text{pre}}$ & $A_{\text{post}}$ & $\Delta A$ & $\rho$ & Cens. \\
\midrule
R-core & 24 & 3 & deg & 631 & 0.111 & 0.210 & +0.099 & 0.24 & 0.52 \\
R-core & 24 & 3 & r$^2$ & 631 & 0.163 & 0.243 & +0.080 & 0.26 & 0.52 \\
R-core & 24 & 5 & deg & 405 & 0.137 & 0.241 & +0.104 & 0.35 & 0.51 \\
R-core & 24 & 5 & r$^2$ & 405 & 0.187 & 0.275 & +0.087 & 0.38 & 0.51 \\
R-core & 24 & 10 & deg & 131 & 0.122 & 0.289 & +0.167 & 0.57 & 0.50 \\
R-core & 24 & 10 & r$^2$ & 131 & 0.178 & 0.320 & +0.142 & 0.59 & 0.50 \\
R-core & 36 & 3 & deg & 631 & 0.103 & 0.081 & -0.023 & 0.14 & 0.92 \\
R-core & 36 & 3 & r$^2$ & 631 & 0.158 & 0.123 & -0.035 & 0.16 & 0.92 \\
R-core & 36 & 5 & deg & 405 & 0.132 & 0.071 & -0.061 & 0.19 & 0.92 \\
R-core & 36 & 5 & r$^2$ & 405 & 0.180 & 0.122 & -0.059 & 0.25 & 0.92 \\
R-core & 36 & 10 & deg & 131 & 0.113 & 0.009 & -0.104 & 0.45 & 0.92 \\
R-core & 36 & 10 & r$^2$ & 131 & 0.154 & 0.087 & -0.067 & 0.47 & 0.92 \\
R-lead & 24 & 3 & deg & 566 & 0.191 & 0.344 & +0.152 & 0.27 & 0.52 \\
R-lead & 24 & 3 & r$^2$ & 566 & 0.252 & 0.388 & +0.137 & 0.30 & 0.52 \\
R-lead & 24 & 5 & deg & 366 & 0.194 & 0.326 & +0.132 & 0.35 & 0.51 \\
R-lead & 24 & 5 & r$^2$ & 366 & 0.252 & 0.370 & +0.117 & 0.40 & 0.51 \\
R-lead & 24 & 10 & deg & 119 & 0.153 & 0.311 & +0.158 & 0.55 & 0.50 \\
R-lead & 24 & 10 & r$^2$ & 119 & 0.211 & 0.346 & +0.134 & 0.57 & 0.50 \\
R-lead & 36 & 3 & deg & 566 & 0.162 & 0.007 & -0.154 & 0.18 & 0.92 \\
R-lead & 36 & 3 & r$^2$ & 566 & 0.233 & 0.140 & -0.092 & 0.21 & 0.92 \\
R-lead & 36 & 5 & deg & 366 & 0.155 & 0.048 & -0.107 & 0.25 & 0.92 \\
R-lead & 36 & 5 & r$^2$ & 366 & 0.222 & 0.119 & -0.103 & 0.30 & 0.92 \\
R-lead & 36 & 10 & deg & 119 & 0.120 & 0.000 & -0.120 & 0.52 & 0.92 \\
R-lead & 36 & 10 & r$^2$ & 119 & 0.156 & 0.071 & -0.085 & 0.54 & 0.92 \\
R-share & 24 & 3 & deg & 632 & 0.229 & 0.371 & +0.143 & 0.24 & 0.52 \\
R-share & 24 & 3 & r$^2$ & 632 & 0.294 & 0.427 & +0.133 & 0.26 & 0.52 \\
R-share & 24 & 5 & deg & 412 & 0.225 & 0.352 & +0.127 & 0.33 & 0.51 \\
R-share\textbf{*} & 24 & 5 & r$^2$ & 412 & 0.291 & 0.403 & +0.112 & 0.36 & 0.51 \\
R-share & 24 & 10 & deg & 134 & 0.176 & 0.322 & +0.146 & 0.53 & 0.51 \\
R-share & 24 & 10 & r$^2$ & 134 & 0.245 & 0.357 & +0.112 & 0.56 & 0.51 \\
R-share & 36 & 3 & deg & 632 & 0.217 & 0.146 & -0.070 & 0.17 & 0.92 \\
R-share & 36 & 3 & r$^2$ & 632 & 0.283 & 0.281 & -0.002 & 0.19 & 0.92 \\
R-share & 36 & 5 & deg & 412 & 0.211 & 0.129 & -0.082 & 0.28 & 0.92 \\
R-share & 36 & 5 & r$^2$ & 412 & 0.277 & 0.271 & -0.006 & 0.31 & 0.92 \\
R-share & 36 & 10 & deg & 134 & 0.153 & 0.032 & -0.121 & 0.49 & 0.92 \\
R-share & 36 & 10 & r$^2$ & 134 & 0.209 & 0.164 & -0.045 & 0.50 & 0.92 \\
\bottomrule
\end{tabular}
\end{table*}

\begin{table*}[p]
\centering
\scriptsize
\setlength{\tabcolsep}{4.5pt}
\caption{Placebo divide, discovery half (all 36 pre/post conditions; * marks the pre-registered
primary pre/post condition). $\rho$ is the Spearman rank correlation of per-coder effects across the divide, and Cens.\ the share of post-era rows right-censored at the estimator stage.
Censoring is 0 by construction, since the placebo pseudo-post window is fully covered.}
\label{tab:a3b}
\begin{tabular}{llllrccccc}
\toprule
Credit & $W$ & Floor & Proxy & $n$ & $A_{\text{pre}}$ & $A_{\text{post}}$ & $\Delta A$ & $\rho$ & Cens. \\
\midrule
R-core & 24 & 3 & deg & 275 & 0.030 & 0.110 & +0.080 & 0.30 & 0.00 \\
R-core & 24 & 3 & r$^2$ & 275 & 0.158 & 0.222 & +0.063 & 0.33 & 0.00 \\
R-core & 24 & 5 & deg & 211 & 0.029 & 0.107 & +0.078 & 0.30 & 0.00 \\
R-core & 24 & 5 & r$^2$ & 211 & 0.158 & 0.218 & +0.059 & 0.32 & 0.00 \\
R-core & 24 & 10 & deg & 91 & 0.006 & 0.079 & +0.073 & 0.34 & 0.00 \\
R-core & 24 & 10 & r$^2$ & 91 & 0.071 & 0.091 & +0.021 & 0.38 & 0.00 \\
R-core & 36 & 3 & deg & 275 & 0.124 & 0.120 & -0.004 & 0.37 & 0.00 \\
R-core & 36 & 3 & r$^2$ & 275 & 0.169 & 0.225 & +0.056 & 0.40 & 0.00 \\
R-core & 36 & 5 & deg & 211 & 0.125 & 0.121 & -0.004 & 0.40 & 0.00 \\
R-core & 36 & 5 & r$^2$ & 211 & 0.170 & 0.226 & +0.056 & 0.42 & 0.00 \\
R-core & 36 & 10 & deg & 91 & 0.018 & 0.089 & +0.071 & 0.45 & 0.00 \\
R-core & 36 & 10 & r$^2$ & 91 & 0.105 & 0.151 & +0.046 & 0.50 & 0.00 \\
R-lead & 24 & 3 & deg & 233 & 0.074 & 0.161 & +0.087 & 0.36 & 0.00 \\
R-lead & 24 & 3 & r$^2$ & 233 & 0.213 & 0.304 & +0.091 & 0.40 & 0.00 \\
R-lead & 24 & 5 & deg & 179 & 0.133 & 0.159 & +0.026 & 0.40 & 0.00 \\
R-lead & 24 & 5 & r$^2$ & 179 & 0.216 & 0.296 & +0.080 & 0.42 & 0.00 \\
R-lead & 24 & 10 & deg & 80 & 0.100 & 0.096 & -0.004 & 0.49 & 0.00 \\
R-lead & 24 & 10 & r$^2$ & 80 & 0.212 & 0.195 & -0.017 & 0.51 & 0.00 \\
R-lead & 36 & 3 & deg & 233 & 0.122 & 0.175 & +0.052 & 0.42 & 0.00 \\
R-lead & 36 & 3 & r$^2$ & 233 & 0.233 & 0.312 & +0.078 & 0.45 & 0.00 \\
R-lead & 36 & 5 & deg & 179 & 0.164 & 0.175 & +0.011 & 0.45 & 0.00 \\
R-lead & 36 & 5 & r$^2$ & 179 & 0.234 & 0.308 & +0.074 & 0.47 & 0.00 \\
R-lead & 36 & 10 & deg & 80 & 0.124 & 0.111 & -0.013 & 0.56 & 0.00 \\
R-lead & 36 & 10 & r$^2$ & 80 & 0.220 & 0.201 & -0.019 & 0.58 & 0.00 \\
R-share & 24 & 3 & deg & 267 & 0.111 & 0.215 & +0.105 & 0.42 & 0.00 \\
R-share & 24 & 3 & r$^2$ & 267 & 0.254 & 0.363 & +0.110 & 0.45 & 0.00 \\
R-share & 24 & 5 & deg & 199 & 0.086 & 0.201 & +0.115 & 0.48 & 0.00 \\
R-share\textbf{*} & 24 & 5 & r$^2$ & 199 & 0.238 & 0.342 & +0.105 & 0.49 & 0.00 \\
R-share & 24 & 10 & deg & 90 & 0.029 & 0.123 & +0.094 & 0.45 & 0.00 \\
R-share & 24 & 10 & r$^2$ & 90 & 0.206 & 0.138 & -0.067 & 0.48 & 0.00 \\
R-share & 36 & 3 & deg & 267 & 0.181 & 0.237 & +0.056 & 0.47 & 0.00 \\
R-share & 36 & 3 & r$^2$ & 267 & 0.268 & 0.380 & +0.112 & 0.49 & 0.00 \\
R-share & 36 & 5 & deg & 199 & 0.182 & 0.225 & +0.043 & 0.53 & 0.00 \\
R-share & 36 & 5 & r$^2$ & 199 & 0.256 & 0.363 & +0.107 & 0.55 & 0.00 \\
R-share & 36 & 10 & deg & 90 & 0.091 & 0.145 & +0.055 & 0.54 & 0.00 \\
R-share & 36 & 10 & r$^2$ & 90 & 0.210 & 0.230 & +0.019 & 0.58 & 0.00 \\
\bottomrule
\end{tabular}
\end{table*}

\begin{table*}[p]
\centering
\scriptsize
\setlength{\tabcolsep}{4.5pt}
\caption{Real divide, held-out half (all 36 pre/post conditions; * marks the pre-registered
primary pre/post condition). $\rho$ is the Spearman rank correlation of per-coder effects across the divide, and Cens.\ the share of post-era rows right-censored at the estimator stage.
Real-divide $W$=36 conditions carry heavy censoring and are reported as a diagnostic, not an estimate (main text).}
\label{tab:a4}
\begin{tabular}{llllrccccc}
\toprule
Credit & $W$ & Floor & Proxy & $n$ & $A_{\text{pre}}$ & $A_{\text{post}}$ & $\Delta A$ & $\rho$ & Cens. \\
\midrule
R-core & 24 & 3 & deg & 652 & 0.156 & 0.255 & +0.099 & 0.23 & 0.52 \\
R-core & 24 & 3 & r$^2$ & 652 & 0.261 & 0.367 & +0.106 & 0.24 & 0.52 \\
R-core & 24 & 5 & deg & 413 & 0.181 & 0.304 & +0.123 & 0.23 & 0.52 \\
R-core & 24 & 5 & r$^2$ & 413 & 0.294 & 0.431 & +0.137 & 0.25 & 0.52 \\
R-core & 24 & 10 & deg & 144 & 0.258 & 0.306 & +0.048 & 0.38 & 0.49 \\
R-core & 24 & 10 & r$^2$ & 144 & 0.374 & 0.416 & +0.042 & 0.43 & 0.49 \\
R-core & 36 & 3 & deg & 652 & 0.153 & 0.036 & -0.117 & 0.26 & 0.93 \\
R-core & 36 & 3 & r$^2$ & 652 & 0.251 & 0.421 & +0.170 & 0.27 & 0.93 \\
R-core & 36 & 5 & deg & 413 & 0.176 & 0.129 & -0.046 & 0.34 & 0.93 \\
R-core & 36 & 5 & r$^2$ & 413 & 0.281 & 0.600 & +0.319 & 0.36 & 0.93 \\
R-core & 36 & 10 & deg & 144 & 0.253 & 0.001 & -0.252 & 0.40 & 0.92 \\
R-core & 36 & 10 & r$^2$ & 144 & 0.363 & 0.296 & -0.068 & 0.44 & 0.92 \\
R-lead & 24 & 3 & deg & 578 & 0.186 & 0.289 & +0.103 & 0.27 & 0.52 \\
R-lead & 24 & 3 & r$^2$ & 578 & 0.295 & 0.418 & +0.122 & 0.29 & 0.52 \\
R-lead & 24 & 5 & deg & 373 & 0.212 & 0.309 & +0.097 & 0.28 & 0.52 \\
R-lead & 24 & 5 & r$^2$ & 373 & 0.322 & 0.441 & +0.119 & 0.31 & 0.52 \\
R-lead & 24 & 10 & deg & 136 & 0.222 & 0.295 & +0.073 & 0.45 & 0.50 \\
R-lead & 24 & 10 & r$^2$ & 136 & 0.376 & 0.412 & +0.036 & 0.48 & 0.50 \\
R-lead & 36 & 3 & deg & 578 & 0.183 & 0.000 & -0.183 & 0.29 & 0.93 \\
R-lead & 36 & 3 & r$^2$ & 578 & 0.292 & 0.048 & -0.244 & 0.34 & 0.93 \\
R-lead & 36 & 5 & deg & 373 & 0.207 & 0.013 & -0.194 & 0.36 & 0.93 \\
R-lead & 36 & 5 & r$^2$ & 373 & 0.319 & 0.008 & -0.310 & 0.41 & 0.93 \\
R-lead & 36 & 10 & deg & 136 & 0.262 & 0.000 & -0.262 & 0.46 & 0.92 \\
R-lead & 36 & 10 & r$^2$ & 136 & 0.370 & 0.000 & -0.370 & 0.53 & 0.92 \\
R-share & 24 & 3 & deg & 650 & 0.196 & 0.278 & +0.082 & 0.26 & 0.52 \\
R-share & 24 & 3 & r$^2$ & 650 & 0.296 & 0.375 & +0.079 & 0.27 & 0.52 \\
R-share & 24 & 5 & deg & 421 & 0.201 & 0.285 & +0.084 & 0.29 & 0.52 \\
R-share\textbf{*} & 24 & 5 & r$^2$ & 421 & 0.294 & 0.381 & +0.088 & 0.30 & 0.52 \\
R-share & 24 & 10 & deg & 143 & 0.229 & 0.255 & +0.025 & 0.44 & 0.49 \\
R-share & 24 & 10 & r$^2$ & 143 & 0.317 & 0.329 & +0.012 & 0.48 & 0.49 \\
R-share & 36 & 3 & deg & 650 & 0.192 & 0.000 & -0.192 & 0.23 & 0.92 \\
R-share & 36 & 3 & r$^2$ & 650 & 0.288 & 0.336 & +0.048 & 0.28 & 0.92 \\
R-share & 36 & 5 & deg & 421 & 0.195 & 0.001 & -0.195 & 0.32 & 0.92 \\
R-share & 36 & 5 & r$^2$ & 421 & 0.289 & 0.283 & -0.006 & 0.37 & 0.92 \\
R-share & 36 & 10 & deg & 143 & 0.221 & 0.000 & -0.221 & 0.46 & 0.92 \\
R-share & 36 & 10 & r$^2$ & 143 & 0.312 & 0.016 & -0.296 & 0.53 & 0.92 \\
\bottomrule
\end{tabular}
\end{table*}

\begin{table*}[p]
\centering
\scriptsize
\setlength{\tabcolsep}{4.5pt}
\caption{Placebo divide, held-out half (all 36 pre/post conditions; * marks the pre-registered
primary pre/post condition). $\rho$ is the Spearman rank correlation of per-coder effects across the divide, and Cens.\ the share of post-era rows right-censored at the estimator stage.
Censoring is 0 by construction, since the placebo pseudo-post window is fully covered.}
\label{tab:a4b}
\begin{tabular}{llllrccccc}
\toprule
Credit & $W$ & Floor & Proxy & $n$ & $A_{\text{pre}}$ & $A_{\text{post}}$ & $\Delta A$ & $\rho$ & Cens. \\
\midrule
R-core & 24 & 3 & deg & 285 & 0.201 & 0.164 & -0.037 & 0.32 & 0.00 \\
R-core & 24 & 3 & r$^2$ & 285 & 0.244 & 0.277 & +0.033 & 0.36 & 0.00 \\
R-core & 24 & 5 & deg & 213 & 0.202 & 0.159 & -0.044 & 0.31 & 0.00 \\
R-core & 24 & 5 & r$^2$ & 213 & 0.249 & 0.279 & +0.029 & 0.35 & 0.00 \\
R-core & 24 & 10 & deg & 102 & 0.245 & 0.176 & -0.069 & 0.49 & 0.00 \\
R-core & 24 & 10 & r$^2$ & 102 & 0.289 & 0.270 & -0.020 & 0.54 & 0.00 \\
R-core & 36 & 3 & deg & 285 & 0.187 & 0.156 & -0.031 & 0.33 & 0.00 \\
R-core & 36 & 3 & r$^2$ & 285 & 0.241 & 0.272 & +0.031 & 0.37 & 0.00 \\
R-core & 36 & 5 & deg & 213 & 0.194 & 0.161 & -0.033 & 0.36 & 0.00 \\
R-core & 36 & 5 & r$^2$ & 213 & 0.248 & 0.281 & +0.033 & 0.40 & 0.00 \\
R-core & 36 & 10 & deg & 102 & 0.229 & 0.172 & -0.057 & 0.45 & 0.00 \\
R-core & 36 & 10 & r$^2$ & 102 & 0.284 & 0.265 & -0.019 & 0.51 & 0.00 \\
R-lead & 24 & 3 & deg & 241 & 0.172 & 0.206 & +0.033 & 0.29 & 0.00 \\
R-lead & 24 & 3 & r$^2$ & 241 & 0.231 & 0.329 & +0.099 & 0.33 & 0.00 \\
R-lead & 24 & 5 & deg & 187 & 0.188 & 0.217 & +0.029 & 0.30 & 0.00 \\
R-lead & 24 & 5 & r$^2$ & 187 & 0.246 & 0.334 & +0.088 & 0.33 & 0.00 \\
R-lead & 24 & 10 & deg & 91 & 0.223 & 0.218 & -0.005 & 0.40 & 0.00 \\
R-lead & 24 & 10 & r$^2$ & 91 & 0.269 & 0.265 & -0.004 & 0.46 & 0.00 \\
R-lead & 36 & 3 & deg & 241 & 0.193 & 0.214 & +0.021 & 0.36 & 0.00 \\
R-lead & 36 & 3 & r$^2$ & 241 & 0.267 & 0.347 & +0.080 & 0.40 & 0.00 \\
R-lead & 36 & 5 & deg & 187 & 0.219 & 0.217 & -0.001 & 0.37 & 0.00 \\
R-lead & 36 & 5 & r$^2$ & 187 & 0.284 & 0.349 & +0.065 & 0.40 & 0.00 \\
R-lead & 36 & 10 & deg & 91 & 0.256 & 0.219 & -0.037 & 0.43 & 0.00 \\
R-lead & 36 & 10 & r$^2$ & 91 & 0.301 & 0.313 & +0.012 & 0.46 & 0.00 \\
R-share & 24 & 3 & deg & 285 & 0.251 & 0.228 & -0.023 & 0.36 & 0.00 \\
R-share & 24 & 3 & r$^2$ & 285 & 0.304 & 0.377 & +0.072 & 0.40 & 0.00 \\
R-share & 24 & 5 & deg & 217 & 0.263 & 0.221 & -0.042 & 0.37 & 0.00 \\
R-share\textbf{*} & 24 & 5 & r$^2$ & 217 & 0.306 & 0.366 & +0.060 & 0.40 & 0.00 \\
R-share & 24 & 10 & deg & 100 & 0.275 & 0.217 & -0.058 & 0.40 & 0.00 \\
R-share & 24 & 10 & r$^2$ & 100 & 0.314 & 0.352 & +0.037 & 0.44 & 0.00 \\
R-share & 36 & 3 & deg & 285 & 0.288 & 0.228 & -0.060 & 0.40 & 0.00 \\
R-share & 36 & 3 & r$^2$ & 285 & 0.337 & 0.387 & +0.050 & 0.44 & 0.00 \\
R-share & 36 & 5 & deg & 217 & 0.299 & 0.220 & -0.079 & 0.44 & 0.00 \\
R-share & 36 & 5 & r$^2$ & 217 & 0.341 & 0.376 & +0.035 & 0.47 & 0.00 \\
R-share & 36 & 10 & deg & 100 & 0.310 & 0.214 & -0.096 & 0.41 & 0.00 \\
R-share & 36 & 10 & r$^2$ & 100 & 0.348 & 0.363 & +0.014 & 0.45 & 0.00 \\
\bottomrule
\end{tabular}
\end{table*}

\begin{table*}[p]
\centering
\scriptsize
\caption{Clean-cadence sensitivity, discovery half (careers beginning August 2017 or later;
$W$=60 horizons fully covered by dense snapshots; 18 conditions). All conditions reject the
lottery null under BH-FDR; the full stable-factor verdict passes in 15/18.}
\label{tab:a5}
\begin{tabular}{llllrccccc}
\toprule
Credit & $W$ & Floor & Proxy & $n$ & TVD & $A$ & 95\% CI & Ratio & Verdict \\
\midrule
R-core & 60 & 3 & deg & 6,404 & 0.029 & 0.113 & [0.07, 0.14] & 0.47 & T \\
R-core & 60 & 3 & r$^2$ & 6,404 & 0.027 & 0.166 & [0.11, 0.23] & 0.12 & F \\
R-core & 60 & 5 & deg & 3,749 & 0.041 & 0.123 & [0.08, 0.16] & 0.55 & T \\
R-core & 60 & 5 & r$^2$ & 3,749 & 0.041 & 0.178 & [0.11, 0.24] & 0.11 & F \\
R-core & 60 & 10 & deg & 1,211 & 0.062 & 0.136 & [0.05, 0.18] & 0.58 & T \\
R-core & 60 & 10 & r$^2$ & 1,211 & 0.056 & 0.208 & [0.10, 0.30] & 0.69 & T \\
R-lead & 60 & 3 & deg & 5,406 & 0.027 & 0.183 & [0.13, 0.21] & 0.61 & T \\
R-lead & 60 & 3 & r$^2$ & 5,406 & 0.031 & 0.232 & [0.18, 0.28] & 0.24 & T \\
R-lead & 60 & 5 & deg & 3,205 & 0.035 & 0.184 & [0.14, 0.22] & 0.65 & T \\
R-lead & 60 & 5 & r$^2$ & 3,205 & 0.048 & 0.233 & [0.18, 0.29] & 0.44 & T \\
R-lead & 60 & 10 & deg & 1,060 & 0.061 & 0.183 & [0.10, 0.22] & 0.62 & T \\
R-lead & 60 & 10 & r$^2$ & 1,060 & 0.059 & 0.233 & [0.11, 0.30] & 0.68 & T \\
R-share & 60 & 3 & deg & 6,284 & 0.024 & 0.262 & [0.21, 0.32] & 1.00 & T \\
R-share & 60 & 3 & r$^2$ & 6,284 & 0.032 & 0.330 & [0.26, 0.41] & 1.00 & F \\
R-share & 60 & 5 & deg & 3,721 & 0.041 & 0.244 & [0.19, 0.29] & 0.24 & T \\
R-share & 60 & 5 & r$^2$ & 3,721 & 0.048 & 0.310 & [0.24, 0.38] & 1.00 & T \\
R-share & 60 & 10 & deg & 1,206 & 0.065 & 0.233 & [0.16, 0.27] & 0.68 & T \\
R-share & 60 & 10 & r$^2$ & 1,206 & 0.077 & 0.303 & [0.20, 0.35] & 0.74 & T \\
\bottomrule
\end{tabular}
\end{table*}

\end{document}